\pdfoutput=1

\documentclass[12pt,a4paper]{article}

\usepackage{ifthen} 
\newboolean{pdflatex}
\setboolean{pdflatex}{true} 

\newboolean{articletitles}
\setboolean{articletitles}{true} 

\newboolean{uprightparticles}
\setboolean{uprightparticles}{false} 

\def\paperauthors{LHCb collaboration} 
\def\paperasciititle{Resolution of outstanding puzzles in B+ -> K+ pi+ pi- decays} 
\def\papertitle{Resolution of outstanding puzzles in $B^+ \!\to K^+ \pi^+ \pi^-$ decays} 
\def\paperkeywords{{High Energy Physics}, {LHCb}} 
\def\papercopyright{\the\year\ CERN for the benefit of the LHCb collaboration} 
\def\paperlicence{CC BY 4.0 licence}
\def\paperlicenceurl{https://creativecommons.org/licenses/by/4.0/}

\newif\ifEnableSectionTOCLinks
\EnableSectionTOCLinksfalse 

\usepackage[top=1in, bottom=1.25in, left=1in, right=1in]{geometry}

\usepackage{microtype}
\usepackage{lineno}  
\usepackage{xspace} 
\usepackage{caption} 

\usepackage{graphicx}  
\usepackage{color}
\usepackage{colortbl}
\graphicspath{{./figs/}} 

\usepackage{amsmath} 
\usepackage{amssymb}
\usepackage{amsfonts}
\usepackage{upgreek} 

\usepackage[normalem]{ulem} 

\newcommand*\patchAmsMathEnvironmentForLineno[1]{%
\expandafter\let\csname old#1\expandafter\endcsname\csname #1\endcsname
\expandafter\let\csname oldend#1\expandafter\endcsname\csname
end#1\endcsname
 \renewenvironment{#1}%
   {\linenomath\csname old#1\endcsname}%
   {\csname oldend#1\endcsname\endlinenomath}%
}
\newcommand*\patchBothAmsMathEnvironmentsForLineno[1]{%
  \patchAmsMathEnvironmentForLineno{#1}%
  \patchAmsMathEnvironmentForLineno{#1*}%
}
\AtBeginDocument{%
\patchBothAmsMathEnvironmentsForLineno{equation}%
\patchBothAmsMathEnvironmentsForLineno{align}%
\patchBothAmsMathEnvironmentsForLineno{flalign}%
\patchBothAmsMathEnvironmentsForLineno{alignat}%
\patchBothAmsMathEnvironmentsForLineno{gather}%
\patchBothAmsMathEnvironmentsForLineno{multline}%
\patchBothAmsMathEnvironmentsForLineno{eqnarray}%
}

\usepackage[pdftex,
            pdfauthor={\paperauthors},
            pdftitle={\paperasciititle},
            pdfkeywords={\paperkeywords}]{hyperref}
\usepackage{hyperxmp}
\hypersetup{
    pdfcopyright={Copyright (C) \papercopyright},
    pdflicenseurl={\paperlicenceurl}
}

\usepackage[colorinlistoftodos,textsize=scriptsize]{todonotes}

\usepackage[bottom,flushmargin,hang,multiple]{footmisc}

\usepackage[all]{hypcap} 

\usepackage{xspace} 
\usepackage{upgreek}

\def\lhcb   {\mbox{LHCb}\xspace}

\def\babar  {\mbox{BaBar}\xspace}
\def\belle  {\mbox{Belle}\xspace}

\def\MagUp {\mbox{\em Mag\kern -0.05em Up}\xspace}

\ifdefined\ifuprightparticles
\else
\newboolean{uprightparticles}
\setboolean{uprightparticles}{false} 
\fi

\ifthenelse{\boolean{uprightparticles}}%
{

 \def\Peta        {\ensuremath{\upeta}\xspace}

 \def\Ppi         {\ensuremath{\uppi}\xspace}

 \def\Ppsi        {\ensuremath{\uppsi}\xspace}

 \def\PDelta      {\ensuremath{\Delta}\xspace}                 
 \def\PXi         {\ensuremath{\Xi}\xspace}                 
 \def\PLambda     {\ensuremath{\Lambda}\xspace}                 
 \def\PSigma      {\ensuremath{\Sigma}\xspace}                 
 \def\POmega      {\ensuremath{\Omega}\xspace}                 
 \def\PUpsilon    {\ensuremath{\Upsilon}\xspace}
 \let\oldPi\Pi
 \def\PPi         {\ensuremath{\oldPi}\xspace}

 \def\PB      {\ensuremath{\mathrm{B}}\xspace}                 
 \def\PD      {\ensuremath{\mathrm{D}}\xspace}                 
 \def\PJ      {\ensuremath{\mathrm{J}}\xspace}                 
 \def\PK      {\ensuremath{\mathrm{K}}\xspace}                 
 \def\Pb      {\ensuremath{\mathrm{b}}\xspace}                 
 \def\Pc      {\ensuremath{\mathrm{c}}\xspace}                 
 \def\Pd      {\ensuremath{\mathrm{d}}\xspace}                 

 \def\Pp      {\ensuremath{\mathrm{p}}\xspace}                 

 \def\Ps      {\ensuremath{\mathrm{s}}\xspace}                 
                  
 \def\Pu      {\ensuremath{\mathrm{u}}\xspace}                 
 \def\thebaroffset{0.0em}
}
{

 \def\Peta        {\ensuremath{\eta}\xspace}

 \def\Ppi         {\ensuremath{\pi}\xspace}

 \def\Ppsi        {\ensuremath{\psi}\xspace}                 
                  
 \mathchardef\PDelta="7101
 \mathchardef\PXi="7104
 \mathchardef\PLambda="7103
 \mathchardef\PSigma="7106
 \mathchardef\POmega="710A
 \mathchardef\PUpsilon="7107
 \mathchardef\PPi="7105
 \def\PB      {\ensuremath{B}\xspace}                 
 \def\PD      {\ensuremath{D}\xspace}                 
 \def\PJ      {\ensuremath{J}\xspace}                 
 \def\PK      {\ensuremath{K}\xspace}                 
 \def\Pb      {\ensuremath{b}\xspace}                 
 \def\Pc      {\ensuremath{c}\xspace}                 
 \def\Pd      {\ensuremath{d}\xspace}                 

 \def\Pp      {\ensuremath{p}\xspace}                 

 \def\Ps      {\ensuremath{s}\xspace}                 
                  
 \def\Pu      {\ensuremath{u}\xspace}                 
 \def\thebaroffset{0.18em}
}
\newcommand{\offsetoverline}[2][\thebaroffset]{\kern #1\overline{\kern -#1 #2}}%

\makeatletter
\ifcase \@ptsize \relax
  \newcommand{\miniscule}{\@setfontsize\miniscule{4}{5}}
\or
  \newcommand{\miniscule}{\@setfontsize\miniscule{5}{6}}
\or
  \newcommand{\miniscule}{\@setfontsize\miniscule{5}{6}}
\fi
\makeatother

\DeclareRobustCommand{\optbar}[1]{\shortstack{{\miniscule (\rule[.5ex]{1.25em}{.18mm})}
  \\ [-.7ex] $#1$}}

\def\uquark    {{\ensuremath{\Pu}}\xspace}

\def\dquark    {{\ensuremath{\Pd}}\xspace}

\def\squark    {{\ensuremath{\Ps}}\xspace}

\def\cquark    {{\ensuremath{\Pc}}\xspace}

\def\bquark    {{\ensuremath{\Pb}}\xspace}

\def\pion   {{\ensuremath{\Ppi}}\xspace}

\def\pip    {{\ensuremath{\pion^+}}\xspace}
\def\pim    {{\ensuremath{\pion^-}}\xspace}

\def\kaon    {{\ensuremath{\PK}}\xspace}
\def\Kbar    {{\ensuremath{\offsetoverline{\PK}}}\xspace}

\def\KorKbar {\kern \thebaroffset\optbar{\kern -\thebaroffset \PK}{}\xspace}
\def\Kz      {{\ensuremath{\kaon^0}}\xspace}

\def\Kp      {{\ensuremath{\kaon^+}}\xspace}

\def\KS      {{\ensuremath{\kaon^0_{\mathrm{S}}}}\xspace}

\newcommand{\etapr}{\ensuremath{\Peta^{\prime}}\xspace}

\def\Dbar    {{\ensuremath{\offsetoverline{\PD}}}\xspace}
\def\D       {{\ensuremath{\PD}}\xspace}

\def\DorDbar {\kern \thebaroffset\optbar{\kern -\thebaroffset \PD}\xspace}

\def\Dzb     {{\ensuremath{\Dbar{}^0}}\xspace}
\def\Dp      {{\ensuremath{\D^+}}\xspace}
\def\Dm      {{\ensuremath{\D^-}}\xspace}

\def\DpDm    {\ensuremath{\Dp {\kern -0.16em \Dm}}\xspace}

\def\B       {{\ensuremath{\PB}}\xspace}
\def\Bbar    {{\ensuremath{\offsetoverline{\PB}}}\xspace}

\def\BorBbar {\kern \thebaroffset\optbar{\kern -\thebaroffset \PB}\xspace}
\def\Bz      {{\ensuremath{\B^0}}\xspace}

\def\Bd      {{\ensuremath{\B^0}}\xspace}

\def\BdorBdbar {\kern \thebaroffset\optbar{\kern -\thebaroffset \Bd}\xspace}
\def\Bu      {{\ensuremath{\B^+}}\xspace}
\def\Bub     {{\ensuremath{\B^-}}\xspace}
\def\Bp      {{\ensuremath{\Bu}}\xspace}
\def\Bm      {{\ensuremath{\Bub}}\xspace}
\def\Bpm     {{\ensuremath{\B^\pm}}\xspace}

\def\Bs      {{\ensuremath{\B^0_\squark}}\xspace}

\def\BsorBsbar {\kern \thebaroffset\optbar{\kern -\thebaroffset \Bs}\xspace}

\def\jpsi     {{\ensuremath{{\PJ\mskip -3mu/\mskip -2mu\Ppsi}}}\xspace}

\def\Y#1S{\ensuremath{\PUpsilon{(#1S)}}\xspace}

\def\proton      {{\ensuremath{\Pp}}\xspace}

\def\LorLbar     {\kern \thebaroffset\optbar{\kern -\thebaroffset \PLambda}\xspace}

\newcommand{\decay}[2]{\ensuremath{\mathinner{#1\!\to #2}}\xspace}

\def\to                 {\ensuremath{\rightarrow}\xspace}

\def\CP                {{\ensuremath{C\!P}}\xspace}

\def\Vud  {{\ensuremath{V_{\uquark\dquark}^{\phantom{\ast}}}}\xspace}
\def\Vcd  {{\ensuremath{V_{\cquark\dquark}^{\phantom{\ast}}}}\xspace}

\def\Vubs  {{\ensuremath{V_{\uquark\bquark}^\ast}}\xspace}
\def\Vcbs  {{\ensuremath{V_{\cquark\bquark}^\ast}}\xspace}

\def\AT#1     {\ensuremath{A_{\mathrm{T}}^{#1}}\xspace}           

\def\C#1      {\ensuremath{\mathcal{C}_{#1}}\xspace}                       
\def\Cp#1     {\ensuremath{\mathcal{C}_{#1}^{'}}\xspace}                    
\def\Ceff#1   {\ensuremath{\mathcal{C}_{#1}^{\mathrm{(eff)}}}\xspace}        
\def\Cpeff#1  {\ensuremath{\mathcal{C}_{#1}^{'\mathrm{(eff)}}}\xspace}       
\def\Ope#1    {\ensuremath{\mathcal{O}_{#1}}\xspace}                       
\def\Opep#1   {\ensuremath{\mathcal{O}_{#1}^{'}}\xspace}                    

\newcommand{\aunit}[1]{\ensuremath{\text{\,#1}}}       

\newcommand{\tev}{\aunit{Te\kern -0.1em V}\xspace}
\newcommand{\gev}{\aunit{Ge\kern -0.1em V}\xspace}
\newcommand{\mev}{\aunit{Me\kern -0.1em V}\xspace}
\newcommand{\kev}{\aunit{ke\kern -0.1em V}\xspace}
\newcommand{\ev}{\aunit{e\kern -0.1em V}\xspace}
 
\newcommand{\mevc}{\ensuremath{\aunit{Me\kern -0.1em V\!/}c}\xspace}
\newcommand{\gevc}{\ensuremath{\aunit{Ge\kern -0.1em V\!/}c}\xspace}
\newcommand{\mevcc}{\ensuremath{\aunit{Me\kern -0.1em V\!/}c^2}\xspace}
\newcommand{\gevcc}{\ensuremath{\aunit{Ge\kern -0.1em V\!/}c^2}\xspace}

\def\fb   {\ensuremath{\aunit{fb}}\xspace}
\def\invfb   {\ensuremath{\fb^{-1}}\xspace}

\def\gsim{{~\raise.15em\hbox{$>$}\kern-.85em
          \lower.35em\hbox{$\sim$}~}\xspace}
\def\lsim{{~\raise.15em\hbox{$<$}\kern-.85em
          \lower.35em\hbox{$\sim$}~}\xspace}

\def\tell1  {TELL1\xspace}
\def\ukl1   {UKL1\xspace}

\newcommand{\lhcborcid}[1]{\href{https://orcid.org/#1}{\hspace*{0.1em}\raisebox{-0.45ex}{\includegraphics[width=1em]{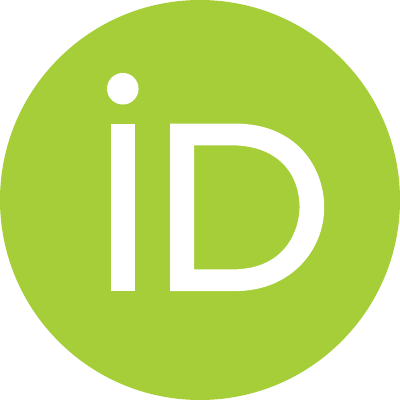}}}}

\hypersetup{
  colorlinks   = true, 
  urlcolor     = blue, 
  linkcolor    = blue, 
  citecolor    = red   
}

\ifEnableSectionTOCLinks
    \usepackage[explicit]{titlesec} 
    
    \let\oldcontentsline\contentsline
    \renewcommand

    \titleformat{\section}{\normalfont\Large\bf}{\hyperlink{tocsection.\thesection}{{\thesection} \parbox[t]{\dimexpr\textwidth-1pc}{#1}}}{1pc}{}

    \titleformat{\subsection}{\normalfont\bf}{\hyperlink{tocsubsection.\thesubsection}{{\thesubsection} \parbox[t]{\dimexpr\textwidth-1pc}{#1}}}{1pc}{}

    \titleformat{name=\section,numberless}[display]{}{}{0pt}{\normalfont\Huge\bfseries #1}
\fi

\usepackage{cite} 
\usepackage{LHCb/mciteplus}
\makeatletter
\g@addto@macro\bfseries{\boldmath}
\makeatother

\usepackage{comment}

\newcommand{\mKpi}{{\ensuremath{m_{\Kp\pim}}}\xspace}
\newcommand{\mpipi}{{\ensuremath{m_{\pip\pim}}}\xspace}

\begin{document}

\renewcommand{\thefootnote}{\fnsymbol{footnote}}
\setcounter{footnote}{1}


\begin{titlepage}
\pagenumbering{roman}

\vspace*{-1.5cm}
\centerline{\large EUROPEAN ORGANIZATION FOR NUCLEAR RESEARCH (CERN)}
\vspace*{1.5cm}
\noindent
\begin{tabular*}{\linewidth}{lc@{\extracolsep{\fill}}r@{\extracolsep{0pt}}}
\ifthenelse{\boolean{pdflatex}}
{\vspace*{-1.5cm}\mbox{\!\!\!\includegraphics[width=.14\textwidth]{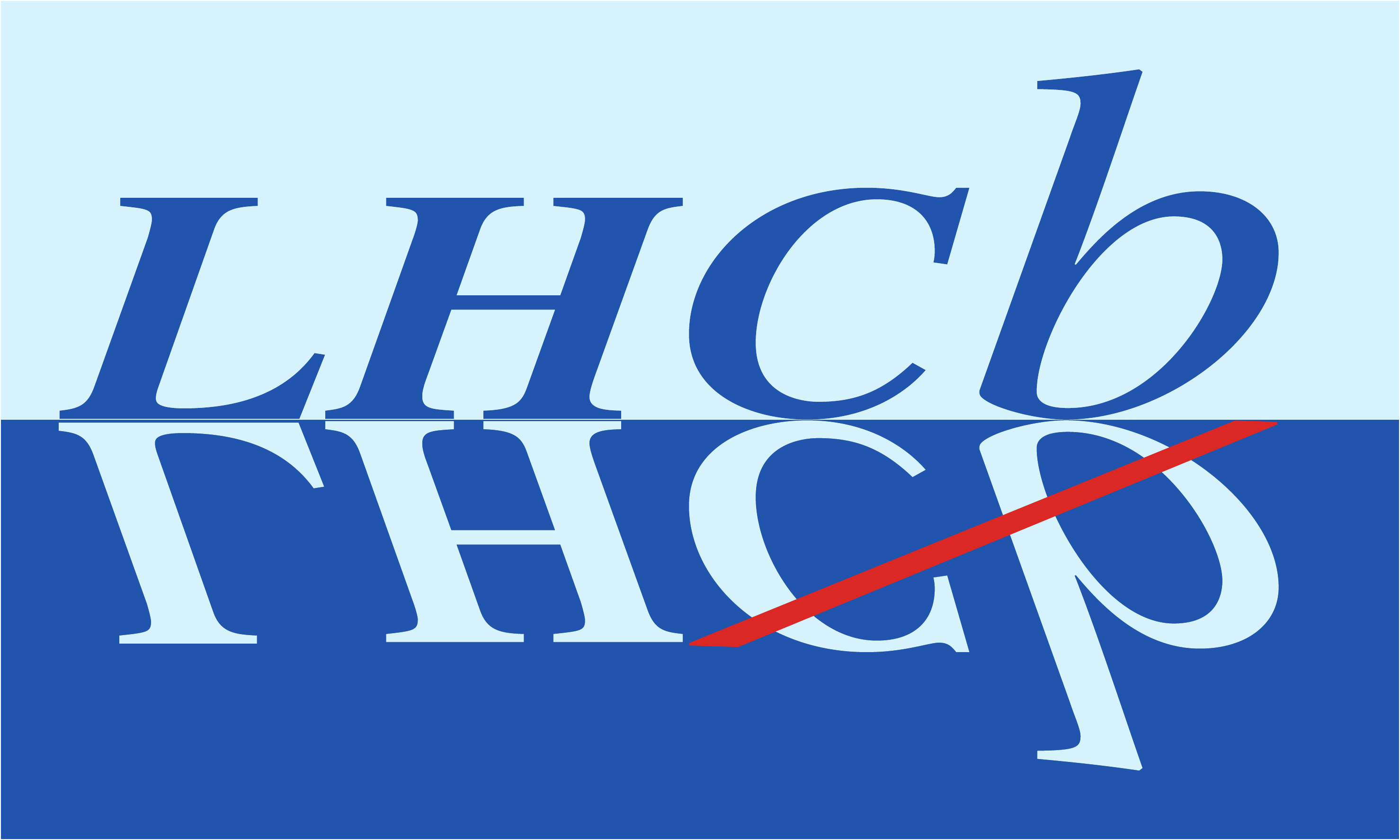}} & &}%
{\vspace*{-1.2cm}\mbox{\!\!\!\includegraphics[width=.12\textwidth]{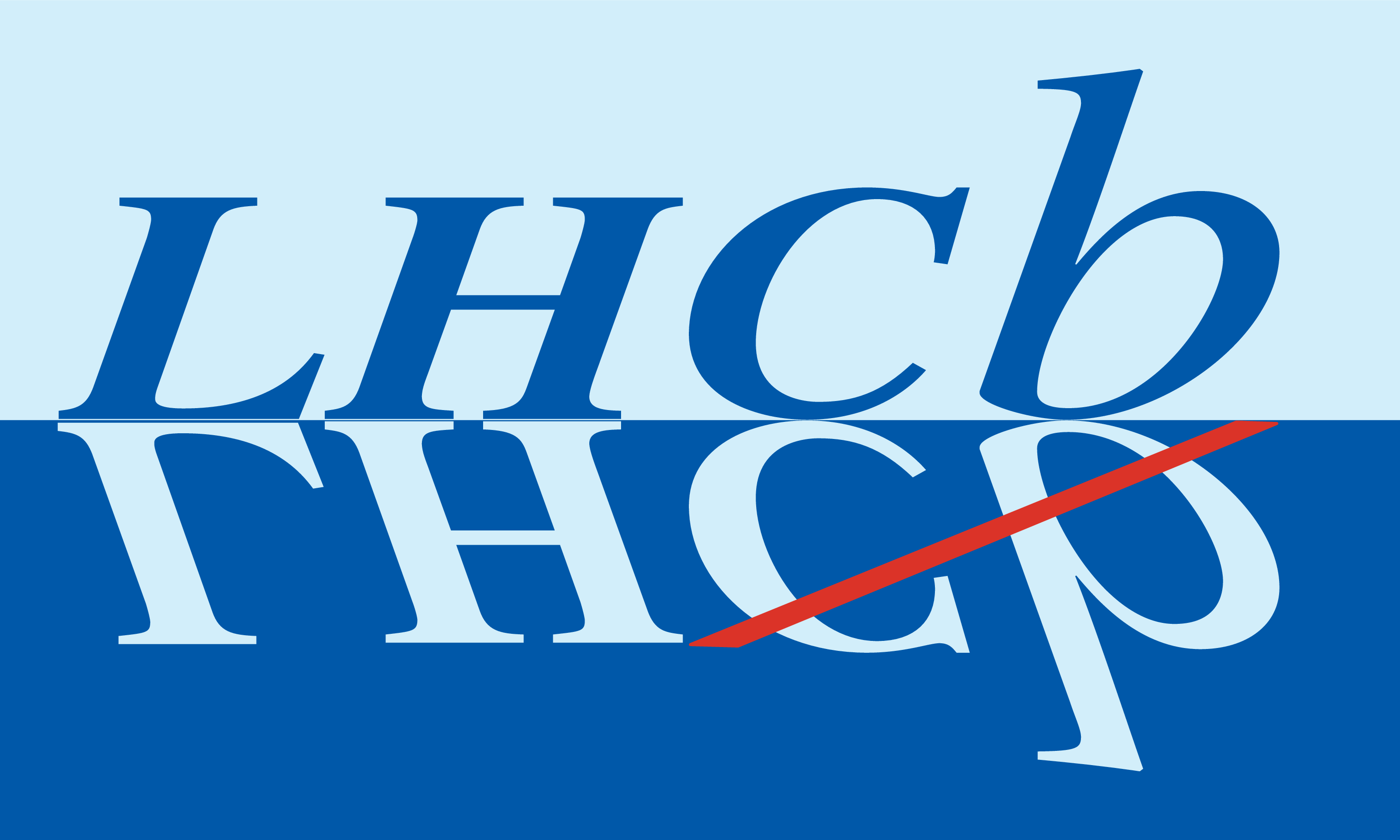}} & &}%
\\
 & & CERN-EP-2026-217 \\  
 & & LHCb-PAPER-2025-068 \\  
 & & 17 August 2026 \\ 
 & & \\
\end{tabular*}

\vspace*{2.0cm}

{\normalfont\bfseries\boldmath\huge
\begin{center}
  \papertitle 
\end{center}
}

\vspace*{1.0cm}

\begin{center}
\paperauthors\footnote{Authors are listed at the end of this Letter.}
\end{center}

\vspace{\fill}

\begin{abstract}
    \noindent
An amplitude analysis of $B^+ \!\to K^+ \pi^+ \pi^-$ decays is presented, based on $pp$ collision data recorded with the LHCb detector in 2011--2012, corresponding to an integrated luminosity of $3\,\text{fb}^{-1}$. Previous studies of the $B \!\to K \pi \pi$ sector have left key unresolved questions concerning the model of the S-wave contributions. A pivotal finding is that relaxing unitarity-based assumptions about the relation between the $K^*_0(1430)^0$ resonance and the slowly varying scalar part in $K^+\pi^-$ leads to considerably better agreement between the model and data. The $B^+ \!\to K^*_0(1430)^0 \pi^+$ branching fraction now challenges the experimental consensus that $B \!\to K^*_0(1430) \pi$ decays dominate the $B \!\to K \pi \pi$ phase space, aligning with the predictions of QCD factorisation rather than perturbative QCD, thus reversing the agreement found in previous measurements. With this increased flexibility, it also becomes possible to model the scalar $\pi^+ \pi^-$ amplitude using established states, eliminating the need for the ad-hoc ``$f_X(1300)$'' component included in previous analyses of the $B \!\to K\pi\pi$ sector. These advances facilitate the discovery of ten intermediate decays.
\end{abstract}

\vspace*{1.0cm}

\begin{center}
  Submitted to
  Phys.~Rev.~Lett.
\end{center}

\vspace{\fill}

{\footnotesize 
\centerline{\copyright~\papercopyright. \href{\paperlicenceurl}{\paperlicence}.}}
\vspace*{2mm}

\end{titlepage}


\newpage
\setcounter{page}{2}
\mbox{~}


\renewcommand{\thefootnote}{\arabic{footnote}}
\setcounter{footnote}{0}


\cleardoublepage


\pagestyle{plain} 
\setcounter{page}{1}
\pagenumbering{arabic}


The challenge of obtaining accurate theoretical predictions within quantum chromodynamics~(QCD) motivates the need for precise experimental measurements to test and refine the relevant theoretical frameworks. 
Charmless hadronic multibody $B$ decays offer an important platform to probe the intricacies of strong-interaction dynamics, because such decays are known to exhibit large matter-antimatter differences~\cite{LHCb-PAPER-2013-027,LHCb-PAPER-2014-044,LHCb-PAPER-2021-049,LHCb-PAPER-2019-018,LHCb-PAPER-2019-017,LHCb-PAPER-2018-051}. 
Otherwise known as charge-parity~(\CP) violation, these effects are driven by weak-interaction phases that are well understood within the Cabibbo–Kobayashi–Maskawa~(CKM) mechanism~\cite{Cabibbo:1963yz,Kobayashi:1973fv}. 
A single amplitude model must simultaneously describe a given \B\ decay and that of its \CP-conjugate \Bbar, with differences in the \B and \Bbar decay densities arising only from the complex coefficients that quantify the coupling to each intermediate component of the amplitude. Thus, the analysis incorporates \CP violation, which in principle allows better disentangling of strong-interaction effects compared to approaches that sum over the \B flavour.

Studies of four separate \decay{\B}{K \pi \pi} decay channels performed have proven particularly challenging, with inconsistent experimental results and no consensus regarding their interpretation. 
This arises in large part from current understanding of the single largest contribution to the total decay rate, \decay{\B}{K^*_0(1430) \pi}~\cite{Belle:2005rpz,BaBar:2008lpx,BaBar:2015pwa,Belle:2006ljg,BaBar:2009jov,BaBar:2011vfx}. 
As summarised in Table~\ref{tab:BFcomparison} for the \decay{\Bp}{K^*_0(1430)^0 \pip} branching fraction in particular, previous measurements performed by the \belle~\cite{Belle:2004drb,Belle:2005rpz} and \babar~\cite{BaBar:2008lpx,BaBar:2015pwa} collaborations disagree on the experimental side. 
The superseded Ref.~\cite{Belle:2004drb} contains the only known record of the disfavoured second solution from \belle. 
The same trend is also seen in related \Bz decays~\cite{Belle:2006ljg,BaBar:2009jov}. 
Furthermore, conflicting theoretical predictions based on QCD factorisation~(QCDf)~\cite{Cheng:2013fba,Cheng:2016shb} and perturbative QCD~(pQCD)~\cite{Shen:2006ms} add another element to the tension, 
with the experimental results clearly favouring pQCD. 
This is contrary to the situation for the related \decay{B}{K^*_0(1430) [\eta^{(\prime)},\, \rho,\, \omega,\, \phi]} decays, where experimental results instead align closely with QCDf, leaving an unresolved puzzle~\cite{Cheng:2013fba}. 
Three-body QCD approaches~\cite{Krankl:2015fha,Cheng:2016shb,Chen:2002th,Wang:2014ira} have begun to emerge as a subject of study in part due to this theoretical challenge. 
Another open question concerns the nature of the ad-hoc ``\decay{f_X(1300)}{\pip\pim}'' structure included in previous analyses~\cite{Belle:2005rpz,BaBar:2008lpx,Belle:2006ljg,Belle:2008til,BaBar:2009jov}. 
While some of those studies considered multiple spin hypotheses and their superpositions, noting consistency with known resonances such as the $f_0(1370)$ or $f_0(1500)$, a clear baseline identification remained elusive, leading to speculation about exotic states~\cite{Minkowski:2004xf,Wang:2006ria,Klempt:2007cp,Ochs:2013gi}.

\begin{table}[!b]
    \centering
    \caption{\label{tab:BFcomparison}
        Comparison of experimental measurements and theoretical predictions of the \decay{\Bp}{K^*_0(1430)^0 \pip} branching fraction, in units of $10^{-6}$. 
        The total uncertainties are quoted, except for the measurements of this analysis, where the first uncertainty corresponds to the combined statistical and systematic uncertainties on the fit fraction, and the second is from external inputs used to obtain the absolute branching fraction. 
        The result from Ref.~\cite{BaBar:2015pwa} comes from an analysis of $B^+ \to \KS \pi^+ \pi^0$ decays, while all other experimental measurements are from $B^+ \to K^+ \pi^- \pi^+$ decays.
}

    \resizebox{\textwidth}{!}{
    \begin{tabular}{@{\hspace{0.0cm}}c
     c|
     c
     c||
     c|
     c||
     c
     c@{\hspace{0.0cm}}} \hline
     \multicolumn{2}{c|}{\belle} & \multicolumn{2}{c||}{\babar} & QCDf & pQCD & \multicolumn{2}{c}{This analysis}\\ \hline

        Ref.~\cite{Belle:2004drb} & 
        Ref.~\cite{Belle:2005rpz} & 
        Ref.~\cite{BaBar:2008lpx} & 
        Ref.~\cite{BaBar:2015pwa} & 
        Ref.~\cite{Cheng:2013fba} & Ref.~\cite{Shen:2006ms} & 
        Isobar & 
        K-matrix \\ 
        \hline
        \rule{0pt}{1.em}
        $8.3\,^{+4.0}_{-2.6}$ &
        $51.6\,^{+7.2}_{-7.7}$ & 
        $32.0\,^{+10.9}_{-\phantom{0}6.1}$ & 
        $34.6\pm5.7$ & 
        $12.9\,^{+4.6}_{-3.7}$ & 
        $47.6\,^{+11.3}_{-10.1}$ & 
        $7.35 \pm 0.50 \pm 0.81 $ & 
        $6.26 \pm 0.57 \pm 0.69 $\\

        \hline
    \end{tabular}
    }
\end{table}

From the experimental point of view, such problems must be addressed by improving the treatment of the $\pip\pim$ and $\Kp\pim$ S-waves. 
This Letter summarises the results relevant in this context from an amplitude analysis of the \decay{\Bp}{\Kp \pip \pim} decay,\footnote{The inclusion of charge-conjugate processes is implied.} based on \proton\proton collision data recorded with the LHCb detector in 2011--2012, corresponding to an integrated luminosity of 3\invfb. 
Compared with previous \decay{\Bp}{\Kp \pip \pim} amplitude analyses~\cite{Belle:2005rpz,BaBar:2008lpx}, the data sample is increased by a factor of 20, with a marked improvement in signal-to-background ratio from roughly 1:1 to about 8:1.
The analysis therefore allows a much more detailed investigation of the underlying hadron dynamics. 
An extensive description of the experimental procedure along with a comprehensive collection of results can be found in a companion article~\cite{BuToKpPipPimPRD}, whilst implications of the results related to \CP violation are discussed in another companion Letter~\cite{BuToKpPipPimPRLCP}. 

The \lhcb detector is a single-arm forward spectrometer covering the pseudorapidity range $2 < \eta < 5$, described in detail in Refs.~\cite{LHCb-DP-2008-001,LHCb-DP-2014-002}. 
The selection of signal candidates follows procedures established for measurements of the relative branching fractions of inclusive \decay{\Bp}{h^+ h^{\prime+} h^{\prime-}} decays, where $h^{(\prime)}$ is a pion or kaon, based on the same data sample~\cite{LHCb-PAPER-2020-031}. 
Minor adjustments to the selection procedures are made for the amplitude analysis; 
for example, the narrow contribution from \decay{\jpsi}{\pip\pim} decays is removed with a veto.

After application of the selection criteria, the \Bp-candidate mass distribution is modelled to obtain signal and background yields. 
The fit function includes components for signal decays, combinatorial background, four-body partially reconstructed decays and misidentified three-body peaking backgrounds. 
Parameters describing the shapes of the signal and combinatorial background are determined directly from data, while those of the remaining backgrounds are fixed from dedicated simulation samples corrected for differences between data and simulation. 
The signal region in the \Bp candidate mass, defined as three times the signal mass resolution on both sides of the \Bp peak, is retained for the subsequent amplitude analysis, comprising 115\,102 \decay{\Bp}{\Kp\pip\pim} candidates with an estimated purity of $(88.7\pm0.3)$\%.

Given the large number of broad overlapping resonances and coupled-channel thresholds, it is particularly challenging to obtain satisfactory phenomenological models for the \pip\pim and \Kp\pim S-waves.
Therefore, three complementary approaches describing the total S-wave amplitude, $A_{\rm S}^\pm$ for \Bpm decays, are considered. 
In each case, a conventional ``isobaric'' paradigm for nonzero-spin resonances is used, in which the factorisation of two-body dynamics from three-body unitarity is the central assumption~\cite{isobar1,isobar2,isobar3}. 
The first model~(Isobar) applies the same approach to the S-wave, which comprises the coherent sum of directly produced intermediate scalar states. 
These include a generalised LASS~(GLASS) function~\cite{Estabrooks:1978de,Aston:1987ir,BaBar:2008inr} describing the \Kp\pim S-wave, together with $\pi\pi$--$\pi\pi$~\cite{Garcia-Martin:2011iqs} and $\pi\pi$--$ K\Kbar$~\cite{Pelaez:2020gnd} scattering terms, both of which are 
multiplied by a form factor accounting for the production of the two-body system in the $B$-meson decay~\cite{AlvarengaNogueira:2015wpj,Bediaga:2013ela}. 
The second replaces the \pip\pim S-wave model with the K-matrix formalism, which enforces two-body unitarity with parameters obtained from scattering data~\cite{Aitchison:1972ay,Chung:1995dx,Anisovich:2002ij,BaBar:2008inr,FOCUS:2003tdy}.
The third implements a quasi-model-independent~(QMI) approach, marking the first instance in which any overlapping crossing partial waves in two different two-body systems produced in particle decay are simultaneously modelled; magnitudes and phases representing individual S-waves are free to vary in intervals~(bins) of the \Kp\pim and \pip\pim masses.

The signal amplitude for \Bp and that for \Bm signal decays is constructed as the sum over $N$ nonzero-spin resonant contributions and the aforementioned S-wave component, \mbox{$A^\pm(\Phi_3) \equiv \sum_{j=1}^N a_j^\pm A_j(\Phi_3) + A_{\rm S}^\pm(\Phi_3)$}, where $\Phi_3$ denotes the candidate position in three-body phase space. 
The $A_j$ terms describe the \CP-conserving dynamical amplitude of resonance $j$, consisting of: the product of a relativistic Breit--Wigner mass propagator or an appropriate variant~\cite{Bugg:2006gc,Flatte:1976xu,Pelaez:2022qby,GS}; the spin-dependent angular distribution~\cite{Zemach:1963bc,Zemach:1968zz}; and Blatt--Weisskopf barrier factors~\cite{Blatt:1952ije,BlattWeisskopf}. 
All established intermediate resonant states~\cite{PDG2020} known to decay either into \Kp\pim or \pip\pim with poles below the \Dzb mass are included in this model. 
The complex coefficients $a_j^\pm \equiv (x_j \pm \Delta x_j) + i(y_j \pm \Delta y_j)$ parametrise the relative contributions of each resonance, where the positive~(negative) signs apply to \Bp~(\Bm) decays. 
The \CP-violating differences between \Bp and \Bm amplitudes are encoded in the freedom of $a_j^\pm$ and the relevant parameters of $A_{\rm S}^\pm$ to take different values for the charge-conjugate processes. 
As there is no sensitivity to the absolute phase of the \Bp and \Bm amplitudes, the coefficients of the largest non-S-wave contribution, the $K^*(892)^0$ resonance, relative to which coefficients of the remaining contributions are measured~(specifically, $x = 1$ and $y = \Delta y = 0$ are fixed for this component). Primary flavour-integrated outputs include the physical intermediate branching fractions $\mathcal{F}_j$, which follow from the charge-dependent quasi-two-body fit fractions. Defined as 
\begin{equation}
    \mathcal{F}^\pm_{j} \equiv \frac{\int |a^\pm_j A_j(\Phi_3)|^2 \textrm{d}\Phi_3}{\int \left(|A^+(\Phi_3)|^2 + |A^-(\Phi_3)|^2\right) \textrm{d}\Phi_3} \,,\ 
    \quad\text{with}\quad \mathcal{F}_{j} = \mathcal{F}^+_{j} + \mathcal{F}^-_{j} \,,
    \label{eq:FFdef}
\end{equation}
their sum over all contributions to the model, both S-wave and nonzero spin, may not equate to 100\% due to interference effects. Also of interest are the \CP-averaged relative phase couplings, $\arg(a_j) = \arg\left((a_j^+ + a_j^-)/2\right)$.

An event likelihood is constructed from the squared magnitude of the signal amplitude modulated by the efficiency variation across the Dalitz plot and convolved with the detector resolution in $m^2_{\pip\pim}$, while background model contributions are derived from data in the \Bp-mass sidebands or calibrated simulation. 
The Isobar and K-matrix approaches are implemented in the \textsc{Laura++} software package~\cite{Laura++}, while a GPU-accelerated fitter based on \textsc{Mint2}~\cite{Mint2} is used for the QMI approach. 

The main sources of experimental systematic uncertainty, detailed in Ref.~\cite{BuToKpPipPimPRD}, originate from: the \B-candidate mass model from which fixed signal and background yields are propagated; the inherent uncertainty in calibration as it pertains to efficiency and background phase-space distributions from simulation further weighted by experimentally determined amplitude models~\cite{LHCb-PAPER-2019-017,LHCb-PAPER-2019-018,LHCb-PAPER-2018-051,BESIII:2017kyd}; and definition of the sideband region in data from which the combinatorial background distribution is taken. 
Systematic uncertainties related to the physical amplitude models are also considered. 
These comprise the variation of fixed masses and widths of established resonances according to their known uncertainties~\cite{PDG2020} and alternative barrier factor radii, in addition to the inclusion of the $K^*_4(2045)^0$ and more speculative resonant structures such as the $f_2(1430)$.

\begin{figure}[!tb]
    \centering
    \includegraphics[width=0.5\linewidth]{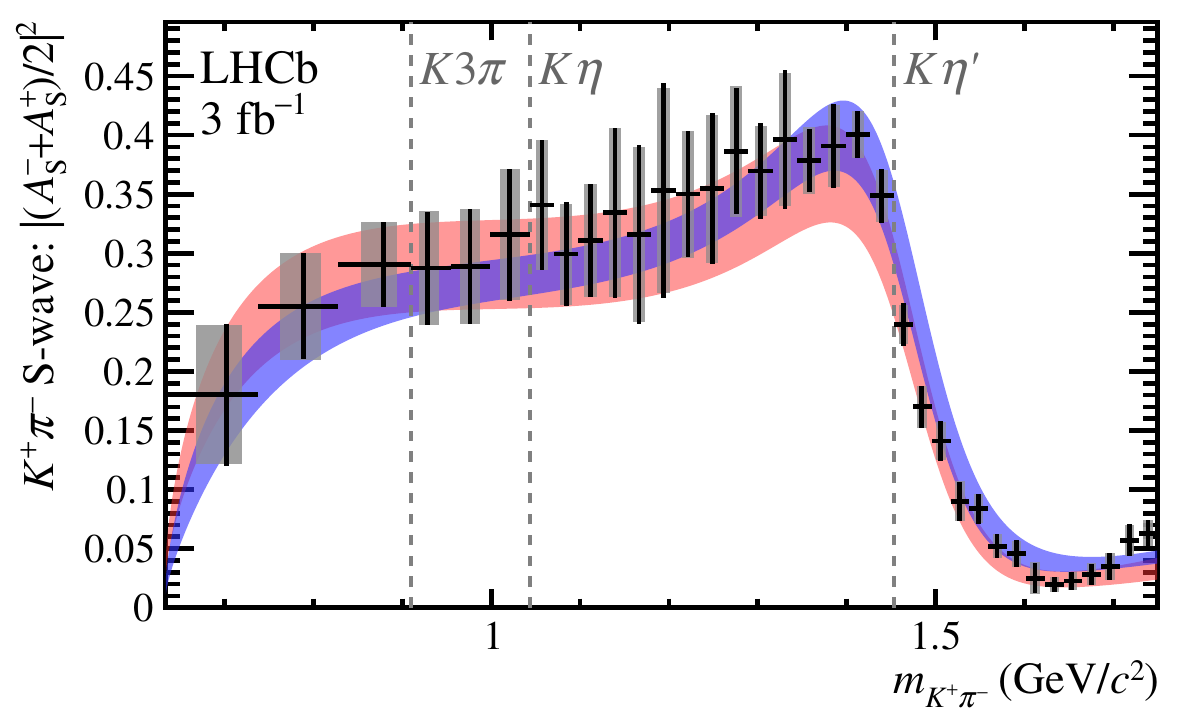}%
    \includegraphics[width=0.5\linewidth]{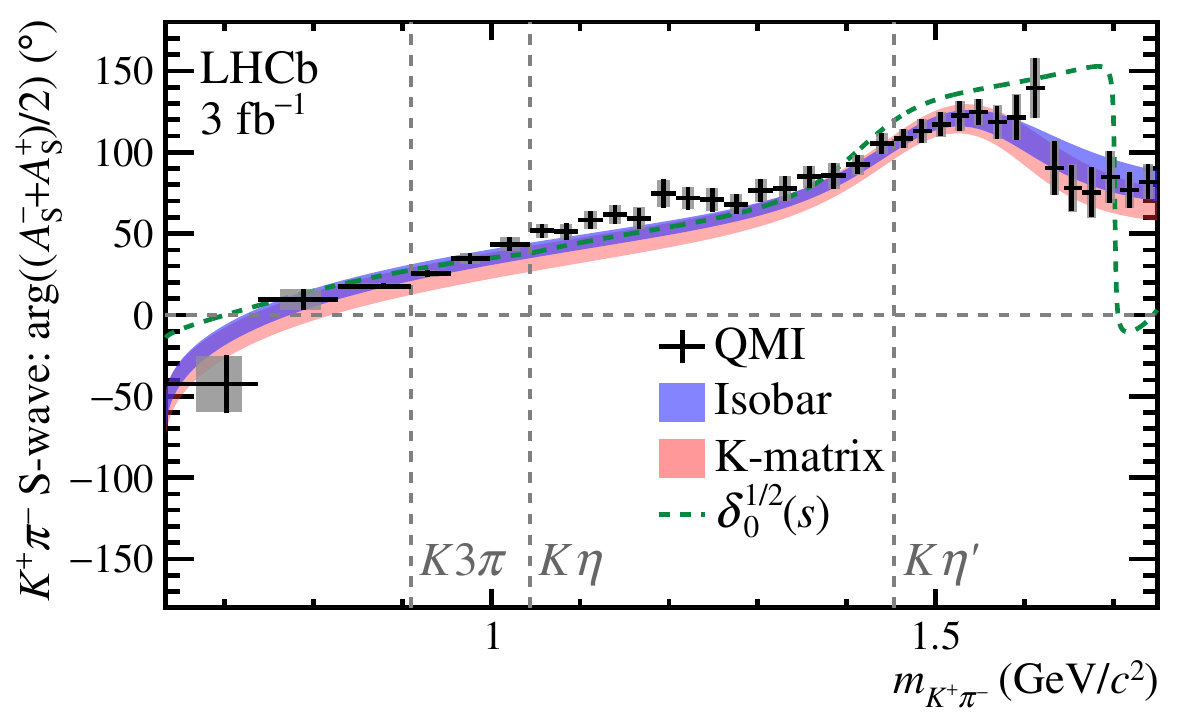}

    \includegraphics[width=0.5\linewidth]{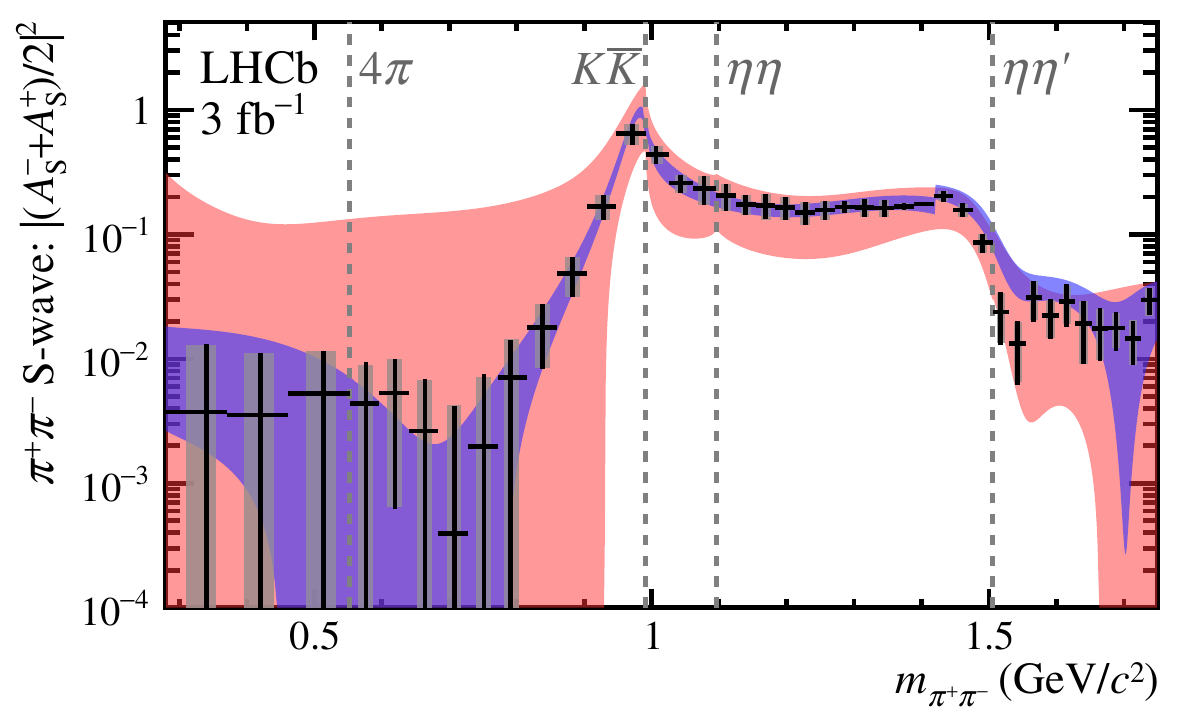}%
    \includegraphics[width=0.5\linewidth]{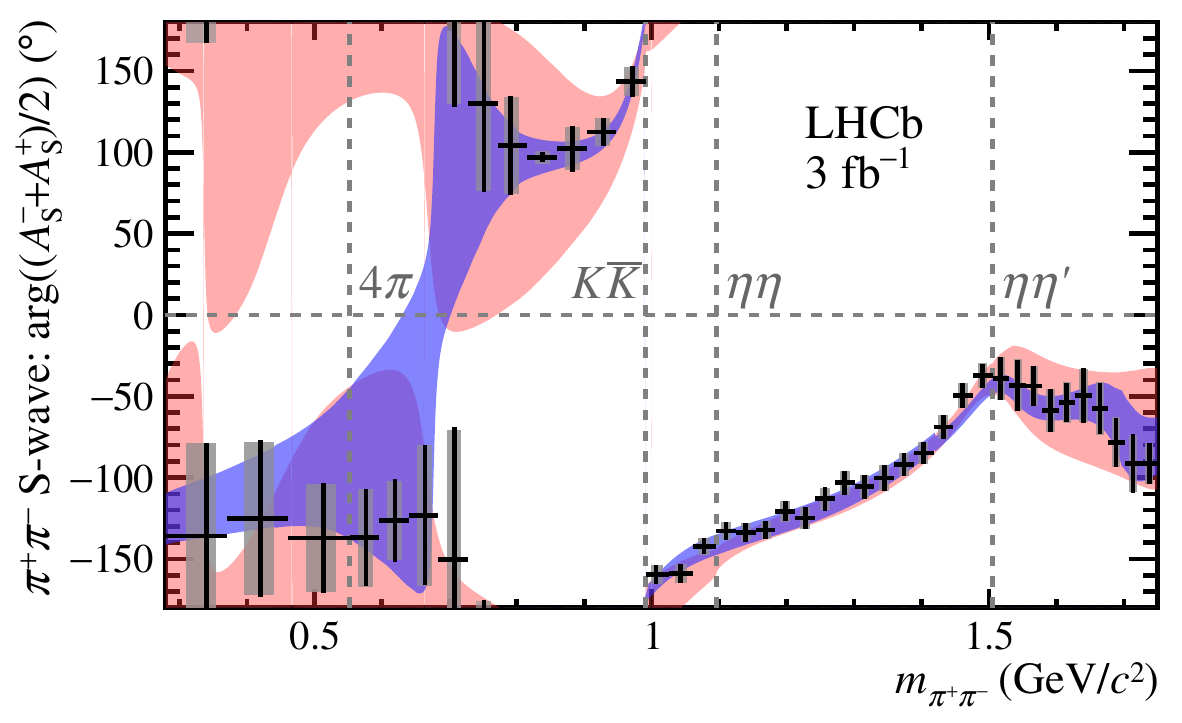}
    \caption{
        Comparison of the S-wave amplitude for (top)~$K^+ \pi^-$ and (bottom)~$\pi^+ \pi^-$, obtained by fitting the LHCb data and including all sources of uncertainty, for the (left) $C\!P$-averaged magnitude squared and (right) $C\!P$-averaged phase in the charmless region. The vertical grey bands show the total QMI uncertainty without accounting for an additional systematic effect arising from the fixed-bin values set from the Isobar result, while this effect is included in the solid black error bars. Vertical dashed lines indicate the openings of coupled channels. 
        The dispersive isospin $I=1/2$ phase of $\pi K$--$\pi K$ scattering $\delta_0^{1/2}(s)$ taken from Ref.~\cite{Pelaez:2020gnd} is arbitrarily offset to overlap the Isobar result and shown for comparison as the green dashed line in the top right plot.
    }
    \label{fig:int:swave}
\end{figure}

Taking into account both statistical and systematic uncertainties, 
good overall agreement between data and the model is obtained for all three S-wave approaches~\cite{BuToKpPipPimPRD}, with the Isobar results showing a better log-likelihood value for nine fewer degrees of freedom compared to the K-matrix approach. On treating the total S-wave as a single contribution in Eq.~\eqref{eq:FFdef}, its total fit fraction as reported in Table~\ref{tab:end:ff} is about 68\% with the results of each approach agreeing within 0.4\%. 
The Isobar and K-matrix S-wave results are projected onto the one-dimensional QMI \mbox{S-waves} in each two-body mass, and compared in the region below the open-charm threshold in Fig.~\ref{fig:int:swave}. 
Plots showing the difference between \Bp and \Bm decays, and extending to the rest of phase space are given in a companion article~\cite{BuToKpPipPimPRD}. 
Due to a complication unique to crossing partial waves~\cite{Krinner:2017dba}, a single QMI bin must be fixed to a reference value, taken from the Isobar model. 
Variation of the amplitude in this bin by the full uncertainty of the Isobar approach is therefore incorporated into the overall QMI systematic uncertainty as an additional effect.

While the magnitude-squared and phases of both the $\Kp \pim$ and $\pip \pim$ S-wave amplitudes of the three S-wave models are generally consistent with each other, a clear discrepancy between the results of the GLASS lineshape and the QMI model can be seen in the \Kp \pim phase above the elastic $K\eta$ threshold and below the $K^*_0(1430)^0$ pole. 
This may motivate future theoretical work to find a suitable analytical replacement for the GLASS lineshape.

The large uncertainties seen in the \mpipi projections for the K-matrix approach are evident in Fig.~\ref{fig:int:swave}. 
In this respect it should be noted that a lineshape obtained from scattering data, in which the $f_0(980)$ pole presents as a slight downwards cusp~\cite{Garcia-Martin:2011iqs}, is being used to model the significant $f_0(980)$ peak in \decay{\Bp}{\Kp \pip \pim} decays. 
The enormous internal interference effects between multiple poles and slowly varying parts, generated in order to produce such a model, are highly sensitive to systematic variations and lead to the large observed uncertainties. 
Conversely, phase-space integrated physical quantities derived from the overall K-matrix S-wave, such as its overall fit fraction and \CP-asymmetry in decay, are found to be substantially more robust to these systematic variations. 
This issue of amplitude stability, along with its poorer log-likelihood value, suggests that the K-matrix approach may not be the most suitable choice for this particular channel with larger data samples. 
The \decay{\Bp}{K^*_0(1430)^0 \pip} contribution forms part of the GLASS function, where the method used to isolate the resonant component and extract its branching fraction is detailed in the companion article~\cite{BuToKpPipPimPRD}.
The measurements of this quantity reported in Table~\ref{tab:BFcomparison} show agreement between the Isobar and K-matrix approaches, but are in stark disagreement with results from previous \decay{\Bp}{\Kp \pip \pim} amplitude analyses~\cite{Belle:2005rpz,BaBar:2008lpx,BaBar:2015pwa} in which the $K^*_0(1430)^0$ fit fractions are between four and eight times larger. 
This occurs since the \Kp\pim S-wave model presented here has some important differences with respect to those used in the previous analyses.
The model used by the \belle\ collaboration~\cite{Belle:2004drb,Belle:2005rpz} consists of an independent Breit--Wigner lineshape for the $K^*_0(1430)^0$ state and an empirical real-valued form for the nonresonant contribution.
This model leads to two solutions that describe the Belle data equally well: one with a high $K^*_0(1430)^0$ fit fraction and an alternative with a low value. 
While comparison with LASS data is used to favour the first solution through invoking Watson's theorem~\cite{Watson:1952ji}, the results of this analysis are in excellent agreement with the disfavoured second solution presented in Ref.~\cite{Belle:2004drb}. 
The \babar\ collaboration analysis, on the other hand, relied on the unitarity-preserving LASS lineshape~\cite{Estabrooks:1978de,Aston:1987ir}, in which the coupling for the term containing the $K^*_0(1430)^0$ resonance is constrained by an effective-range contribution. 
Conversely, the analysis presented here does not impose unitarity preservation, since the GLASS lineshape allows for an independent coupling between its two components. 
The capacity of this lineshape to violate unitarity is justified for several reasons: an additional phase in \decay{\Bp}{(\Kp\pim)\pip} decays compared to that of elastic $K\pi$ scattering could arise from contributions from inelastic $K\eta^{(\prime)}$ coupled channels; an $I=3/2$ amplitude; the interaction between the companion \pip and \Kp\pim S-wave; or the production vertex of the penguin amplitude that is expected to dominate \decay{\Bp}{K_0^{*0}\pip} decays~\cite{Gronau:1994rj}. 

LASS data~\cite{Aston:1987ir} upon which an $I=1/2$ dispersive $\pi K$--$\pi K$ scattering phase is also based~\cite{Pelaez:2020gnd}, exhibit a sharper uptick about the $K^*_0(1430)^0$ pole followed by a sudden drop, suggesting its contribution is considerable in scattering scenarios. 
However, the Isobar and K-matrix results, with support from the QMI approach, indicate a less abrubt variation shown in Fig.~\ref{fig:int:swave}~(top right), meaning impact of the $K^*_0(1430)^0$ resonance is relatively limited. Furthermore, no alternative solution with a high $K^*_0(1430)^0$ fit fraction and reasonable fit quality has been found, nor is the QMI approach able to converge to any solution in support of this outcome. 
That demonstrates that the origin of the large $K^*_0(1430)^0$ fit fraction in previous analyses can be understood as resulting from the presumption of unitarity conservation in an environment where this need not hold. 
The puzzle~\cite{Cheng:2013fba} as to why QCDf does not work well for \decay{B}{K^*_0(1430) \pi} decays, but does for related processes, while the converse is true for pQCD, is thus resolved: the results of this analysis are in full agreement with the predictions of QCDf given in Table~\ref{tab:BFcomparison}. 

\begin{figure}[!tb]
    \centering
    \includegraphics[width=0.5\linewidth]{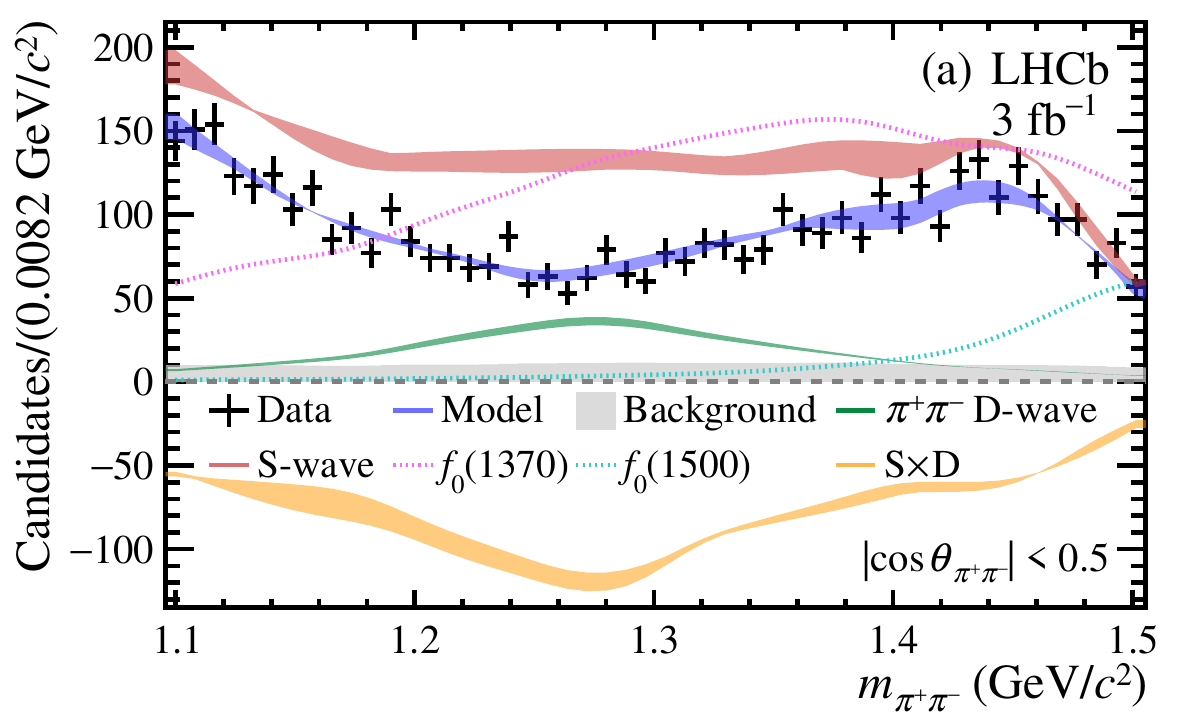}%
    \includegraphics[width=0.5\linewidth]{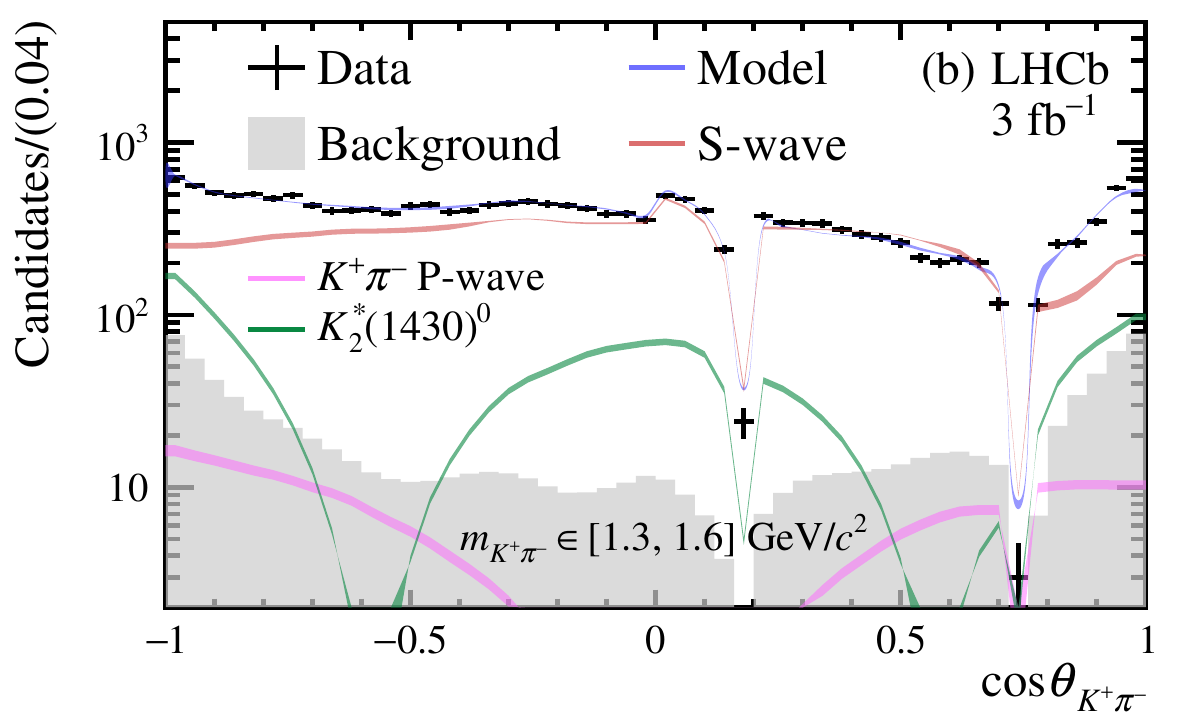}

    \includegraphics[width=0.5\linewidth]{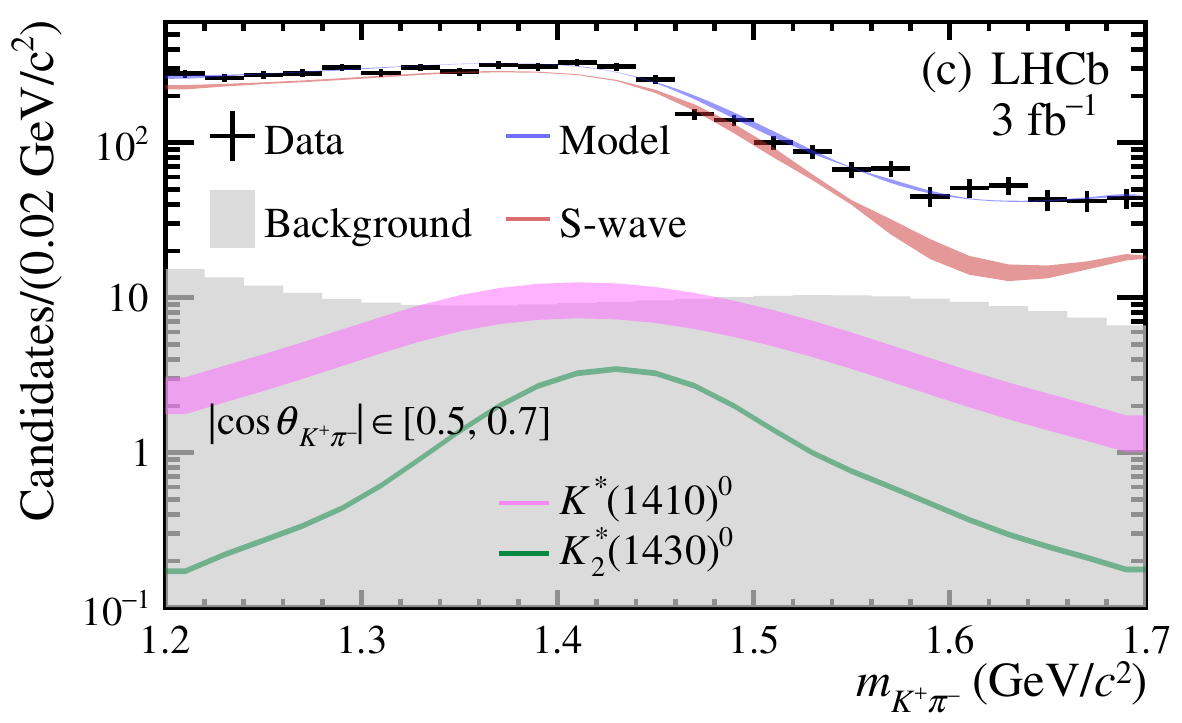}%
    \includegraphics[width=0.5\linewidth]{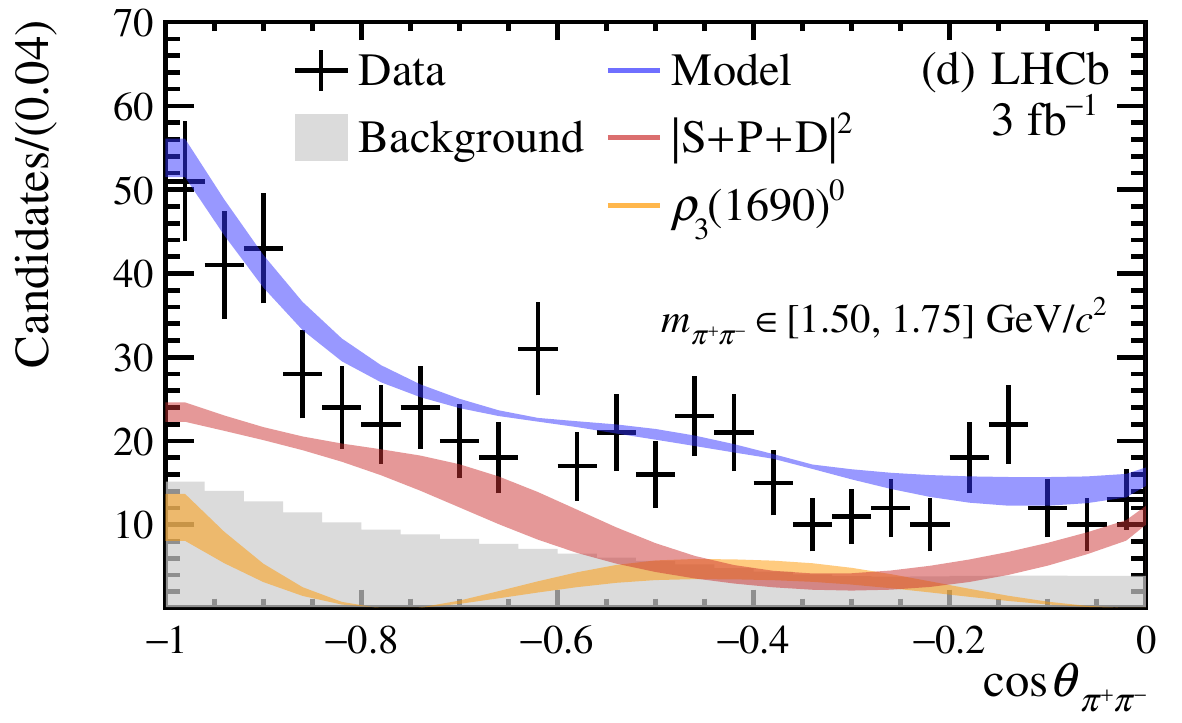}

    \caption{
         Data and model projections enhanced in the (a)~$\pi^+ \pi^-$ inelastic rescattering, (b)~$K_2^*(1430)^0$, (c)~$K^*(1410)^0$ and (d)~$\rho_3(1690)^0$ regions. Bands on the total model and its components indicate the maximum difference between the central values obtained for each S-wave approach, however those unique to the Isobar model are instead indicated by dotted lines. The depleted regions for (b) in $\cos\theta_{K^+ \pi^-}$ arise from the $J\mskip -3mu/\mskip -2mu\psi$ and $\kern 0.18em \overline{\kern -0.18em D}{}^{0}$ mass vetoes.
    }
    \label{fig:int:resonances}
\end{figure}

Moreover, with the improved modelling of the \Kp\pim S-wave implemented in this analysis, a good description of the data in the \mpipi spectrum is achieved in the Isobar approach with the S-wave described by only established states, including the $f_0(1370)$ and $f_0(1500)$ resonances, and the $\pi\pi$ rescattering terms. 
There is no longer any need to include an ad-hoc ``$f_X(1300)$'' $\to \pip\pim$ contribution. 
Similarly, the K-matrix model without any modification to accommodate a new state provides a good description of the data, while the structure of the total S-wave in this \pip\pim mass region is corroborated by the QMI approach.
However, if the \belle or \babar descriptions of the \Kp\pim S-wave~\cite{Belle:2005rpz,BaBar:2008lpx} are used, a model containing a single $f_X$ state with free parameters gives a significantly better log-likelihood value than one including the $f_0(1370)$ and $f_0(1500)$ resonances with fixed parameters. 
Both fits have much worse log-likelihood values than the baseline model with the GLASS lineshape.

The \pip\pim mass distribution between the openings of the $\eta\eta$ and $\eta\eta^\prime$ rescattering channels is shown in Fig.~\ref{fig:int:resonances}(a), where the relative contribution from the S-wave has been enhanced by requiring $|\cos\theta_{\pip\pim}| < 0.5$, which preferentially reduces contributions from \pip\pim states with spin greater than zero and the majority of all other \Kp\pim waves. 
The helicity angle for an intermediate state $\theta_{h^+h^{\prime-}}$ is defined between the momentum of the hadron with opposite charge to the $B$ meson and that of the spectator hadron from the $B$ decay, evaluated in the $h^+h^{\prime-}$ rest frame. 
This region is dominated by the \pip\pim S- and D-waves, with the S-wave comprised of the $f_0(980)$ tail and destructive interference between the $f_0(1370)$ and $f_0(1500)$ intermediate states.
In addition, there is large destructive interference between the S- and D-waves. 

\begin{table}[!tb]
    \centering
    \caption{\label{tab:significance}
        Fit fractions (\%) of intermediate decay channels observed for the first time in each S-wave approach. The first uncertainty is statistical and the second systematic. 
        In the top half, all significance estimates accounting for the total uncertainty and deemed overwhelming~($>\!10\sigma$). 
        This is not the case in the lower half, so their values are presented. 
    }

    \resizebox{\linewidth}{!}{
    \renewcommand{\arraystretch}{1.1}
    \begin{tabular}
    {@{\hspace{0.0cm}}l@{\hspace{0.0cm}}
     @{\hspace{0.125cm}}r@{\hspace{0.0cm}}
     @{\hspace{0.125cm}}r@{\hspace{0.125cm}}
     @{\hspace{0.125cm}}r@{\hspace{0.0cm}}
     @{\hspace{0.125cm}}r@{\hspace{0.125cm}}
     @{\hspace{0.125cm}}r@{\hspace{0.0cm}}
     @{\hspace{0.125cm}}r@{\hspace{0.0cm}}} \hline
    Component 
    & \multicolumn{1}{c}{Isobar} & 
    & \multicolumn{1}{c}{K-matrix} &
    & \multicolumn{1}{c}{QMI} & \\
    \hline

$K_2^*(1430)^0$ & $2.253 \pm 0.125 \pm 0.303$ & & $2.434 \pm 0.117 \pm 0.278$ & & $2.220 \pm 0.119 \pm 0.264$ & \\
$f_2(1270)$ & $2.152 \pm 0.110 \pm 0.278$ & & $2.145 \pm 0.094 \pm 0.240$ & & $2.442 \pm 0.174 \pm 0.406$ & \\
$K_0^*(1950)^0$ & $0.793 \pm 0.092 \pm 0.133$ & & $0.842 \pm 0.079 \pm 0.110$ & & \multicolumn{1}{c}{---} &\\
$f_0(1370)$ & $16.785 \pm 0.951 \pm 1.683$ & & \multicolumn{1}{c}{---} & & \multicolumn{1}{c}{---} &\\
$f_0(1500)$ & $1.857 \pm 0.150 \pm 0.286$ & & \multicolumn{1}{c}{---} & & \multicolumn{1}{c}{---} &\\\hline
$K^*(1410)^0$ & $0.700 \pm 0.092 \pm 0.158$ &$(7.9\sigma)$ & $0.517 \pm 0.088 \pm 0.173$ &$(6.7\sigma)$ & $0.891 \pm 0.166 \pm 0.347$ &$(8.0\sigma)$\\
$\rho(1450)^0$ & $1.213 \pm 0.212 \pm 1.024$ & $(4.0\sigma)$ & $0.881 \pm 0.153 \pm 0.125$ & $(7.4\sigma)$ & $1.908 \pm 0.409 \pm 0.471$ & $(7.4\sigma)$\\
$\rho_3(1690)^0$ & $0.408 \pm 0.047 \pm 0.094$ & $(7.1\sigma)$ & $0.439 \pm 0.042 \pm 0.058$ & $(10.2\sigma)$ & $0.258 \pm 0.043 \pm 0.093$ & $(6.5\sigma)$\\
$f_0(500)$ & $1.833 \pm 0.214 \pm 0.320$ & $(8.2\sigma)$ & \multicolumn{1}{c}{---} & & \multicolumn{1}{c}{---} &\\
$f_0(1710)$ & $0.714 \pm 0.094 \pm 0.588$ & $(5.0\sigma)$ & \multicolumn{1}{c}{---} & & \multicolumn{1}{c}{---} &\\
$\pi\pi$--$\pi\pi$ & $5.365 \pm 0.833 \pm 1.789$ & $(8.5\sigma)$ & \multicolumn{1}{c}{---} & & \multicolumn{1}{c}{---} &\\

    \hline
    \end{tabular}
    }
\end{table}

The significance values of discovered intermediate transitions in units of standard deviations~($\sigma$) are listed in Table~\ref{tab:significance}. 
Ten decay channels are observed beyond $5\sigma$ for the first time based on a more sensitive and statistically robust approach to assessing significance that retains the two-dimensional complex-valued information; technical details are described in Appendix~\ref{sec:appendix:significance}. 
With this improved metric, the first evidence exceeding $3\sigma$ is also obtained for \decay{\Bp}{\rho(1450)^0 \Kp} decays.

The necessity of the intermediate scalar states is demonstrated in the agreement between the Isobar and QMI S-wave amplitudes, as shown in Fig.~\ref{fig:int:swave} and related plots in Ref.~\cite{BuToKpPipPimPRD} showing the difference in S-wave amplitude between \Bp and \Bm decays. 
While the $f_2(1270)$ state is significant, it manifests itself through a largely destructive interference with the S-wave, as presented in Fig.~\ref{fig:int:resonances}(a), so does not present as a peak in $m_{\pip\pim}$. 
By limiting the \mKpi range to $[1.3, 1.6] \gevcc$, the impact of the $K_2^*(1430)^0$ contribution can be seen in the extremes of the helicity regions in Fig.~\ref{fig:int:resonances}(b). 
The $K^*(1410)^0$ state also occupies the same mass range, and as can be seen from the $\cos \theta_{\Kp\pim}$ projection, is larger than the $K_2^*(1430)^0$ contribution in the region $0.5 < |\cos \theta_{\Kp\pim}| < 0.7$. 
This region is selected for the \mKpi projection in Fig.~\ref{fig:int:resonances}(c), making evident the significant $K^*(1410)^0$ contribution. 
Finally, the contribution from the $\rho_3(1690)^0$ state is seen in Fig.~\ref{fig:int:resonances}(d), where $\mpipi \in [1.50, 1.75] \gevcc$ is required. 
Concentrating on the negative helicity angle region where the contributions from \Kp\pim reflections are limited, a notable spin-3 excess above the combined S-, P- and D-waves is seen to be necessary to describe the data, particularly in the $\cos\theta_{\pip\pim} < -0.9$ region.

To conclude, an amplitude analysis of the \decay{\Bp}{\Kp\pip\pim} decay has been performed with \proton\proton collision data recorded with the LHCb detector, corresponding to an integrated luminosity of 3\invfb. Three complementary approaches describe the dominant S-wave contribution, revealing a rich and complex structure that was previously unobserved or unexplained. 
Together with significant \CP-violating effects discussed in companion articles~\cite{BuToKpPipPimPRD,BuToKpPipPimPRLCP}, two long-standing questions related to strong-interaction effects are addressed. 
As regards the S-wave amplitude, the Isobar and K-matrix approaches are mostly in agreement with the QMI results, although a discrepancy in the \Kp\pim phase motion between $K\eta$ and $K\etapr$ thresholds suggests that an improved analytical replacement for the GLASS lineshape will become necessary. 
Nevertheless, the capacity to decouple the $K^*_0(1430)^0$ state from the underlying slowly varying structure is found to challenge the experimental consensus that \decay{B}{K^*_0(1430) \pi} dominates the \decay{B}{K \pi \pi} phase space, and reconcile the \decay{\Bp}{K^*_0(1430)^0 \pip} branching fraction to the predictions of two-body QCDf. 
In addition, questions concerning the ad-hoc ``$f_X(1300)$'' contribution included in previous \decay{B}{K\pi\pi} amplitude analyses are also resolved. 
These advances enable the first observation of ten intermediate decay processes. 

The results presented have considerable intrinsic interest as they provide improved understanding of the underlying strong interactions responsible for the large \CP-violating effects observed across the \decay{\Bp}{h^+ h^{\prime+} h^{\prime-}} phase space.
Moreover, the developments presented here lay the groundwork for their inclusion into the amplitude models of related \B decays, which can be used to search for beyond the Standard Model contributions to loop-mediated multibody charmless hadronic $B$ decays. 
The improved understanding of both $K\pi$ and $\pi\pi$ S-waves will be crucial to search for subtle deviations from the Standard Model predictions for decay-time-dependent \CP violation effects in \decay{\Bz}{\Kz \pip \pim} decays~\cite{Belle:2008til,BaBar:2009jov}. 
The improved modelling will also lead to reduced uncertainties in theoretically clean model-independent isospin sum rules in the \decay{\B}{K^* \pi} and $\rho K$ systems~\cite{Lipkin:1991st,Gronau:2005kz,Gronau:2010dd}, where knowledge of the \decay{\Bp}{K^{*0}\pip} branching fraction is critical since a single decay topology dominates. 
The benefit of such reductions in uncertainty also extends to determinations of the fundamental \CP-violating phase of the CKM matrix, $\gamma = \phi_3 \equiv \arg[-\Vud \Vubs/\Vcd \Vcbs]$, from decays dominated by loop-mediated processes~\cite{Bhattacharya:2013cla,Bhattacharya:2014eca,Bertholet:2018tmx,Bhattacharya:2023pef}.

\section*{Acknowledgements}
%
%
\noindent We express our gratitude to our colleagues in the CERN
accelerator departments for the excellent performance of the LHC. We
thank the technical and administrative staff at the LHCb
institutes.
We acknowledge support from CERN and from the national agencies:
ARC (Australia);
CAPES, CNPq, FAPERJ and FINEP (Brazil); 
MOST and NSFC (China); 
CNRS/IN2P3 and CEA (France);  
BMFTR, DFG and MPG (Germany);
NKFIH (Hungary);              
INFN (Italy); 
NWO (Netherlands); 
MNiSW and NCN (Poland); 
MEC/IFA (Romania); 
MICIU and AEI (Spain);
SNSF and SER (Switzerland); 
NASU (Ukraine); 
STFC (United Kingdom); 
DOE NP and NSF (USA).
We acknowledge the computing resources that are provided by ARDC (Australia), 
CBPF (Brazil),
CERN, 
IHEP and LZU (China),
IN2P3 (France), 
KIT and DESY (Germany), 
INFN (Italy), 
SURF (Netherlands),
Polish WLCG (Poland),
IFIN-HH (Romania), 
PIC (Spain), CSCS (Switzerland), 
GridPP (United Kingdom),
and NSF (USA).  
We are indebted to the communities behind the multiple open-source
software packages on which we depend.
Individual groups or members have received support from
RTP (Australia), 
FWO Odysseus grant G0ASD25N (Belgium), 
Key Research Program of Frontier Sciences of CAS, CAS PIFI, CAS CCEPP (China); 
Minciencias (Colombia);
EPLANET, Marie Sk\l{}odowska-Curie Actions, ERC and NextGenerationEU (European Union);
A*MIDEX, ANR, IPhU and Labex P2IO, and R\'{e}gion Auvergne-Rh\^{o}ne-Alpes (France);
Alexander-von-Humboldt Foundation (Germany);
ICSC (Italy); 
Severo Ochoa and Mar\'ia de Maeztu Units of Excellence, GVA, XuntaGal, GENCAT, InTalent-Inditex and Prog.~Atracci\'on Talento CM (Spain);
the Leverhulme Trust, the Royal Society and UKRI (United Kingdom).

\newpage
\section*{End matter}
\label{sec:appendix}

\appendix

\section{Fit fractions and strong phases}
\label{sec:appendix:FFs-and-phases}

\begin{figure}[tbp]
    \centering
    \includegraphics[width=0.5\linewidth]{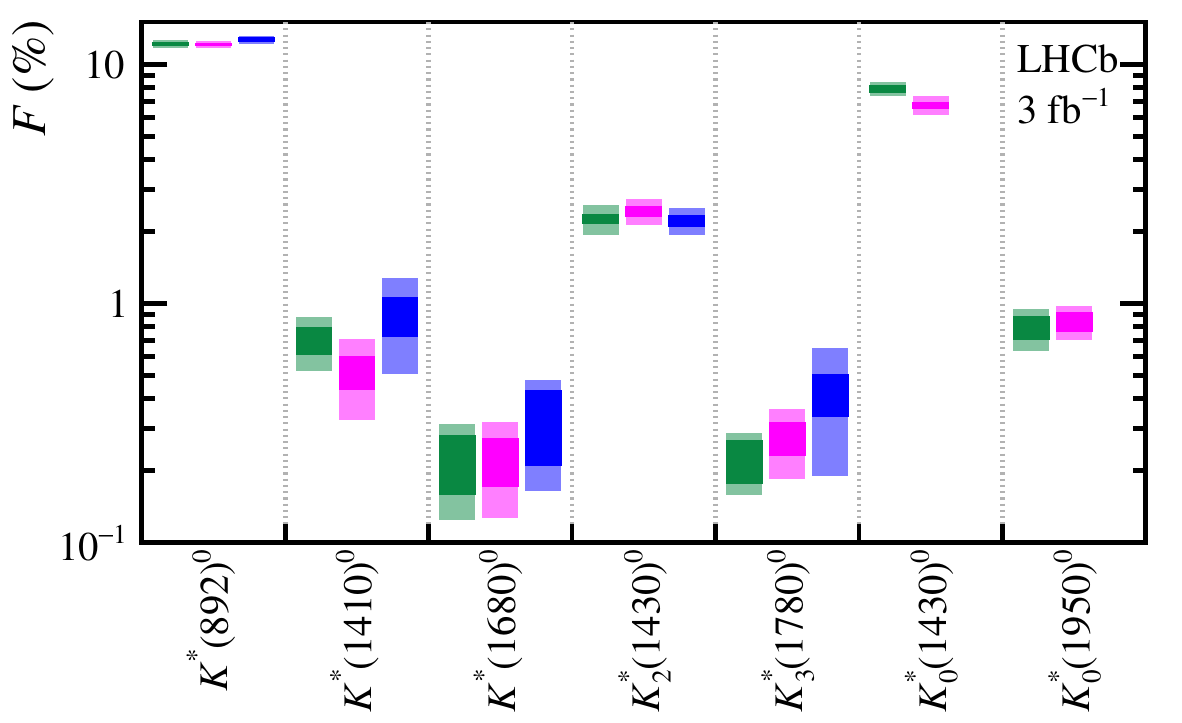}%
    \includegraphics[width=0.5\linewidth]{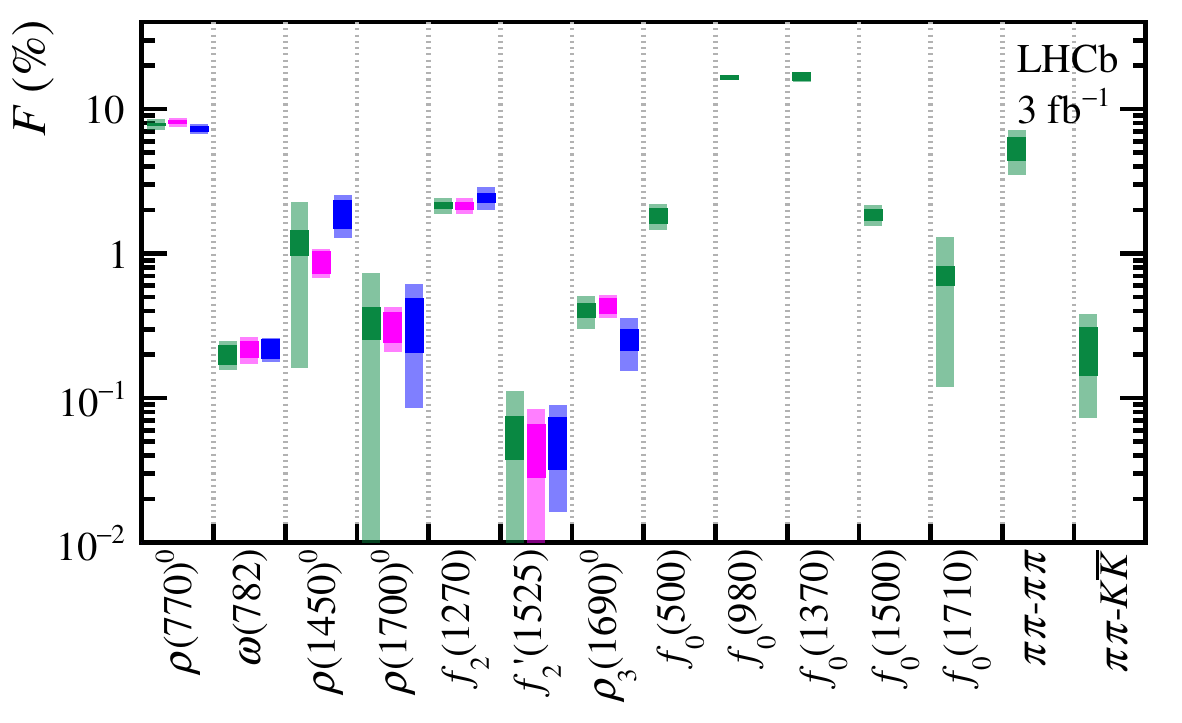}

    \caption{
        Comparison of the quasi-two-body fit fractions between Isobar~(green), K-matrix~(magenta) and QMI~(blue) for states in (left)~$K^+ \pi^-$ and (right)~$\pi^+ \pi^-$. Solid bands show the statistical spread while the lighter shading indicates ranges for the total uncertainty.
    }
    \label{fig:end:ff}
\end{figure}

\begin{figure}[tb]
    \centering
    \includegraphics[width=0.5\linewidth]{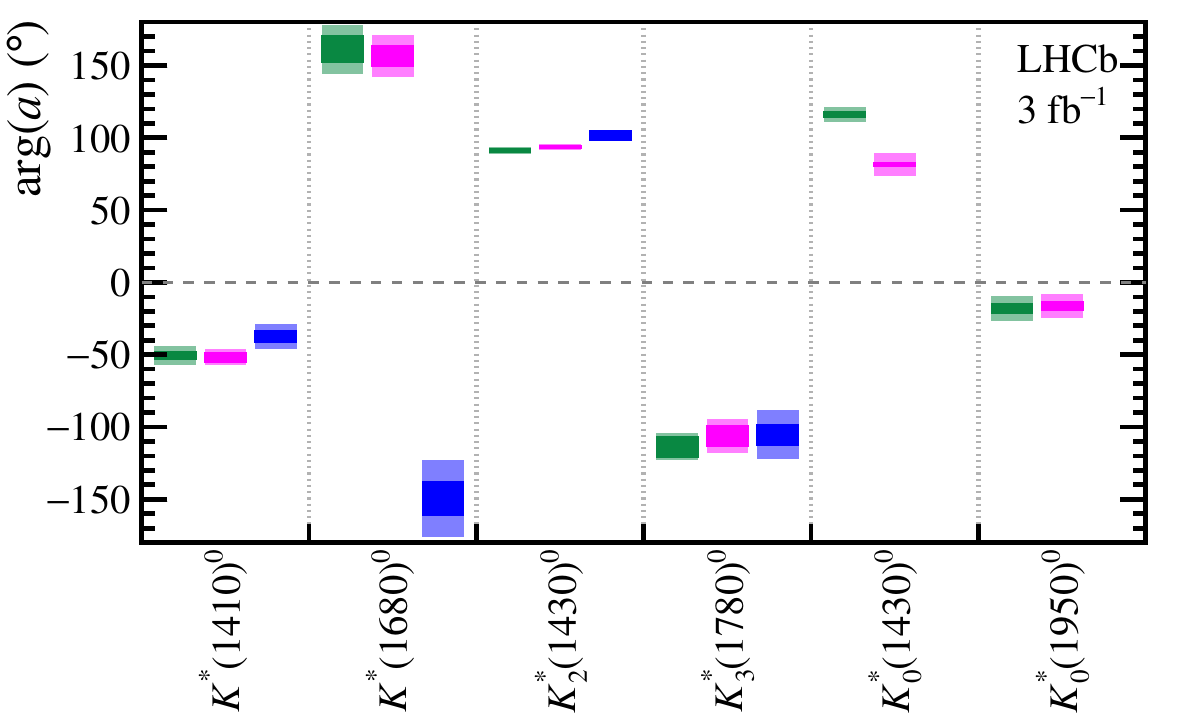}%
    \includegraphics[width=0.5\linewidth]{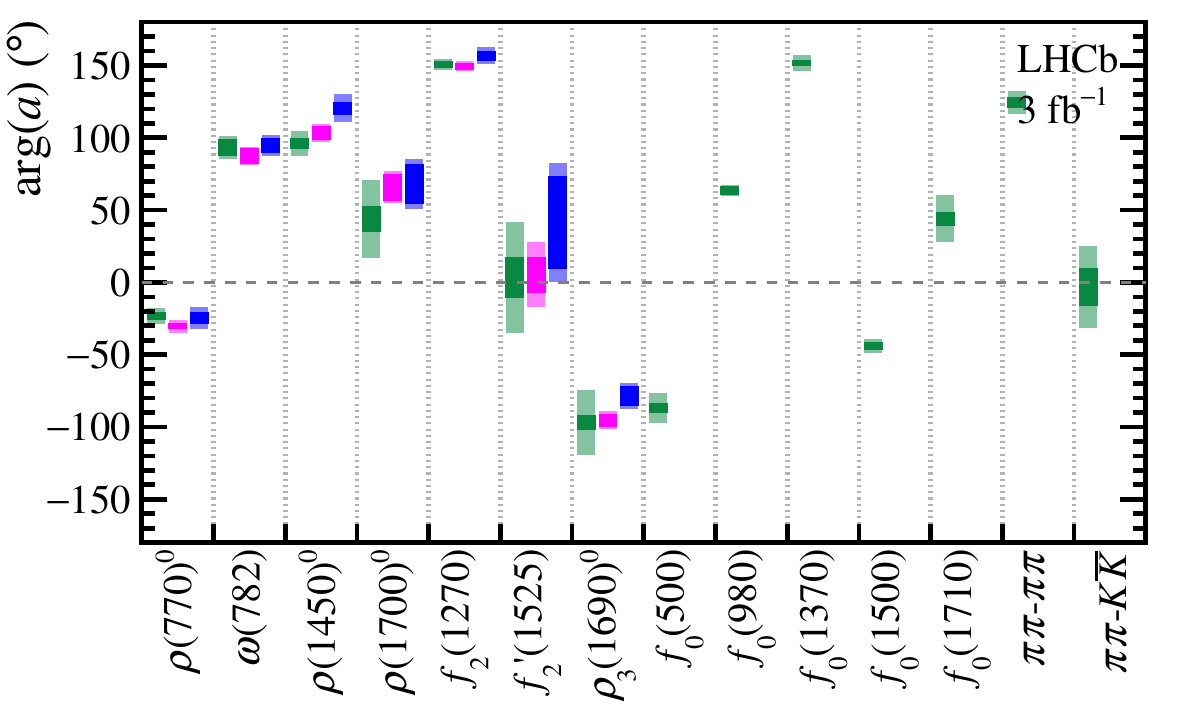}

    \caption{
        Comparison of the average strong phases for each component, relative to the $B^+ \!\to K^*(892)^0 \pi^+$ reference amplitude, between Isobar~(green), K-matrix~(magenta) and QMI~(blue) for states in (left)~$K^+ \pi^-$ and (right)~$\pi^+ \pi^-$. Solid bands show the statistical spread while the lighter shading indicates ranges for the total uncertainty.
    }
    \label{fig:end:phase}
\end{figure}

In this section, the hadronic observables that characterise the amplitude models are presented. The fit fractions for the remaining contributions not already included in Table~\ref{tab:significance} are reported in Table~\ref{tab:end:ff}, where to facilitate a meaningful comparison between S-wave approaches, the total fit fraction is determined from the combined S-wave in \Kp\pim and \pip\pim. 
To ease comparison, all quasi-two-body fit fractions are visually represented in Fig.~\ref{fig:end:ff}. 
The strong phase differences for each component are listed in Table~\ref{tab:end:phase} and shown in Fig.~\ref{fig:end:phase}.

\begin{table}[b]
    \centering
    \caption{\label{tab:end:ff}
        Remaining quasi-two-body fit fractions (\%) not already included in the main text for the three S-wave approaches.
        For all results, the first uncertainty is statistical while the second is systematic. 
    }
    \renewcommand{\arraystretch}{1.1}
    \resizebox{\linewidth}{!}{
    \begin{tabular}
    {@{\hspace{0.0cm}}l@{\hspace{0.125cm}}
     @{\hspace{0.125cm}}c@{\hspace{0.125cm}}
     @{\hspace{0.125cm}}c@{\hspace{0.125cm}}
     @{\hspace{0.125cm}}c@{\hspace{0.0cm}}} \hline
    Component & Isobar & K-matrix & QMI \\ \hline

$K^*(892)^0$ & $12.167 \pm 0.128 \pm 0.518\phantom{1}$ & $12.071 \pm 0.118 \pm 0.400\phantom{1}$ & $12.669 \pm 0.223 \pm 0.417\phantom{1}$\\
$K^*(1680)^0$ & $0.219 \pm 0.060 \pm 0.075$ & $0.223 \pm 0.065 \pm 0.081$ & $0.322 \pm 0.111 \pm 0.113$\\
$\rho(770)^0$ & $7.834 \pm 0.192 \pm 0.652$ & $8.122 \pm 0.194 \pm 0.552$ & $7.326 \pm 0.259 \pm 0.565$\\
$\omega(782)$ & $0.202 \pm 0.030 \pm 0.034$ & $0.219 \pm 0.035 \pm 0.038$ & $0.221 \pm 0.032 \pm 0.027$\\
$\rho(1700)^0$ & $0.340 \pm 0.088 \pm 0.380$ & $0.317 \pm 0.073 \pm 0.078$ & $0.351 \pm 0.142 \pm 0.223$\\
$f_2^\prime(1525)$ & $0.056 \pm 0.019 \pm 0.052$ & $0.047 \pm 0.019 \pm 0.032$ & $0.053 \pm 0.021 \pm 0.030$\\
$K_3^*(1780)^0$ & $0.222 \pm 0.039 \pm 0.044$ & $0.274 \pm 0.042 \pm 0.077$ & $0.420 \pm 0.084 \pm 0.214$\\
$\chi_{c0}(1P)$ & $2.000 \pm 0.065 \pm 0.097$ & $1.927 \pm 0.052 \pm 0.079$ & $1.823 \pm 0.077 \pm 0.146$\\\hline
$K^*_0(1430)^0$ & $7.912 \pm 0.270 \pm 0.465$ & $6.736 \pm 0.208 \pm 0.570$ & ---\\
$f_0(980)$ & $16.582 \pm 0.440 \pm 0.634\phantom{1}$ & --- & ---\\
$\pi\pi$--$K \Kbar$ & $0.227 \pm 0.101 \pm 0.667$ & --- & ---\\\hline
\pip\pim S-wave & --- & $30.901 \pm 0.303 \pm 1.026\phantom{1}$ & --- \\\hline
\Kp\pim and & $68.248 \pm 0.387 \pm 0.788\phantom{1}$ & $68.475 \pm 0.329 \pm 0.432\phantom{1}$ & $67.547 \pm 0.606 \pm 0.641\phantom{1}$ \\
\pip\pim S-waves \\\hline
Total & $98.433 \pm 0.301 \pm 0.386\phantom{1}$ & $98.090 \pm 0.377 \pm 0.339\phantom{1}$ & $98.451 \pm 0.431 \pm 0.567\phantom{1}$\\

    \hline
    \end{tabular}
    }
\end{table}

\begin{table}[tbp]
    \centering
    \caption{\label{tab:end:phase}
        Average strong phases~($^\circ$) for each component, relative to the \decay{\Bp}{K^*(892)^0\pip} reference amplitude, shown for the three S-wave approaches. For all results, the first uncertainty is statistical while the second is systematic. 
    }
    \renewcommand{\arraystretch}{1.1}
    \begin{tabular}
    {@{\hspace{0.0cm}}l@{\hspace{0.125cm}}
     @{\hspace{0.125cm}}c@{\hspace{0.125cm}}
     @{\hspace{0.125cm}}c@{\hspace{0.125cm}}
     @{\hspace{0.125cm}}c@{\hspace{0.0cm}}} \hline
    Component & Isobar & K-matrix & QMI \\ \hline

$K^*(892)^0$ & 0 (fixed) & 0 (fixed) & 0 (fixed) \\
$K^*(1410)^0$ & $\phantom{0}{-50.7} \pm \phantom{0}2.7 \pm \phantom{0}6.0$ & $\phantom{0}{-51.8} \pm \phantom{0}3.4 \pm \phantom{0}4.6$ & $\phantom{0}{-37.4} \pm \phantom{0}4.4 \pm \phantom{0}7.8$\\
$K^*(1680)^0$ & $+161.3 \pm \phantom{0}9.2 \pm 14.3$ & $+156.7 \pm \phantom{0}7.6 \pm 12.7$ & $-149.5 \pm 12.0 \pm 23.6$\\
$\rho(770)^0$ & $\phantom{0}{-23.0} \pm \phantom{0}2.3 \pm \phantom{0}4.9$ & $\phantom{0}{-30.2} \pm \phantom{0}1.8 \pm \phantom{0}4.1$ & $\phantom{0}{-24.4} \pm \phantom{0}3.7 \pm \phantom{0}6.6$\\
$\omega(782)$ & $\phantom{0}{+93.3} \pm \phantom{0}5.3 \pm \phantom{0}5.8$ & $\phantom{0}{+87.7} \pm \phantom{0}5.1 \pm \phantom{0}3.6$ & $\phantom{0}{+94.9} \pm \phantom{0}5.0 \pm \phantom{0}5.4$\\
$\rho(1450)^0$ & $\phantom{0}{+96.0} \pm \phantom{0}3.7 \pm \phantom{0}7.7$ & $+103.4 \pm \phantom{0}4.2 \pm \phantom{0}4.9$ & $+120.6 \pm \phantom{0}4.2 \pm \phantom{0}8.4$\\
$\rho(1700)^0$ & $\phantom{0}{+44.1} \pm \phantom{0}8.7 \pm 25.5$ & $\phantom{0}{+65.7} \pm \phantom{0}9.0 \pm \phantom{0}6.3$ & $\phantom{0}{+68.1} \pm 13.2 \pm 11.6$\\
$K_2^*(1430)^0$ & $\phantom{0}{+91.2} \pm \phantom{0}1.2 \pm \phantom{0}2.2$ & $\phantom{0}{+93.6} \pm \phantom{0}1.1 \pm \phantom{0}1.8$ & $+101.6 \pm \phantom{0}3.2 \pm \phantom{0}2.5$\\
$f_2(1270)$ & $+150.8 \pm \phantom{0}1.9 \pm \phantom{0}3.6$ & $+149.5 \pm \phantom{0}2.1 \pm \phantom{0}2.8$ & $+156.7 \pm \phantom{0}3.0 \pm \phantom{0}5.0$\\
$f_2^\prime(1525)$ & $\phantom{00}{+3.4} \pm 13.7 \pm 36.1$ & $\phantom{00}{+5.3} \pm 12.2 \pm 19.1$ & $\phantom{0}{+41.4} \pm 32.0 \pm 26.0$\\
$K_3^*(1780)^0$ & $-113.5 \pm \phantom{0}7.3 \pm \phantom{0}6.0$ & $-106.1 \pm \phantom{0}7.5 \pm \phantom{0}9.0$ & $-105.3 \pm \phantom{0}7.4 \pm 15.1$\\
$\rho_3(1690)^0$ & $\phantom{0}{-96.9} \pm \phantom{0}4.9 \pm 21.7$ & $\phantom{0}{-95.3} \pm \phantom{0}4.0 \pm \phantom{0}5.0$ & $\phantom{0}{-78.4} \pm \phantom{0}6.5 \pm \phantom{0}6.2$\\
$\chi_{c0}(1P)$ & $+115.4 \pm \phantom{0}2.4 \pm \phantom{0}1.8$ & $+116.1 \pm \phantom{0}2.2 \pm \phantom{0}1.7$ & $+123.0 \pm \phantom{0}3.8 \pm \phantom{0}2.4$\\\hline
$K^*_0(1430)^0$ & $+116.2 \pm \phantom{0}2.4 \pm \phantom{0}5.0$ & $\phantom{0}{+81.6} \pm \phantom{0}1.7 \pm \phantom{0}7.5$ & ---\\
$K_0^*(1950)^0$ & $\phantom{0}{-18.2} \pm \phantom{0}3.5 \pm \phantom{0}7.9$ & $\phantom{0}{-16.5} \pm \phantom{0}3.2 \pm \phantom{0}7.8$ & ---\\\hline
$f_0(500)$ & $\phantom{0}{-86.9} \pm \phantom{0}3.4 \pm \phantom{0}9.7$ & --- & ---\\
$f_0(980)$ & $\phantom{0}{+63.6} \pm \phantom{0}2.5 \pm \phantom{0}3.3$ & --- & ---\\
$f_0(1370)$ & $+151.9 \pm \phantom{0}1.9 \pm \phantom{0}5.1$ & --- & ---\\
$f_0(1500)$ & $\phantom{0}{-43.7} \pm \phantom{0}2.5 \pm \phantom{0}4.1$ & --- & ---\\
$f_0(1710)$ & $\phantom{0}{+44.0} \pm \phantom{0}4.5 \pm 15.6$ & --- & ---\\
$\pi\pi$--$\pi\pi$ & $+124.4 \pm \phantom{0}3.6 \pm \phantom{0}7.3$ & --- & ---\\
$\pi\pi$--$KK$ & $\phantom{00}{-3.2} \pm 12.6 \pm 25.6$ & --- & ---\\\hline

    \hline
    \end{tabular}
\end{table}

\section{Significance of complex couplings}
\label{sec:appendix:significance}

\begin{figure}[!tb]
    \centering
    \includegraphics[width=0.5\linewidth]{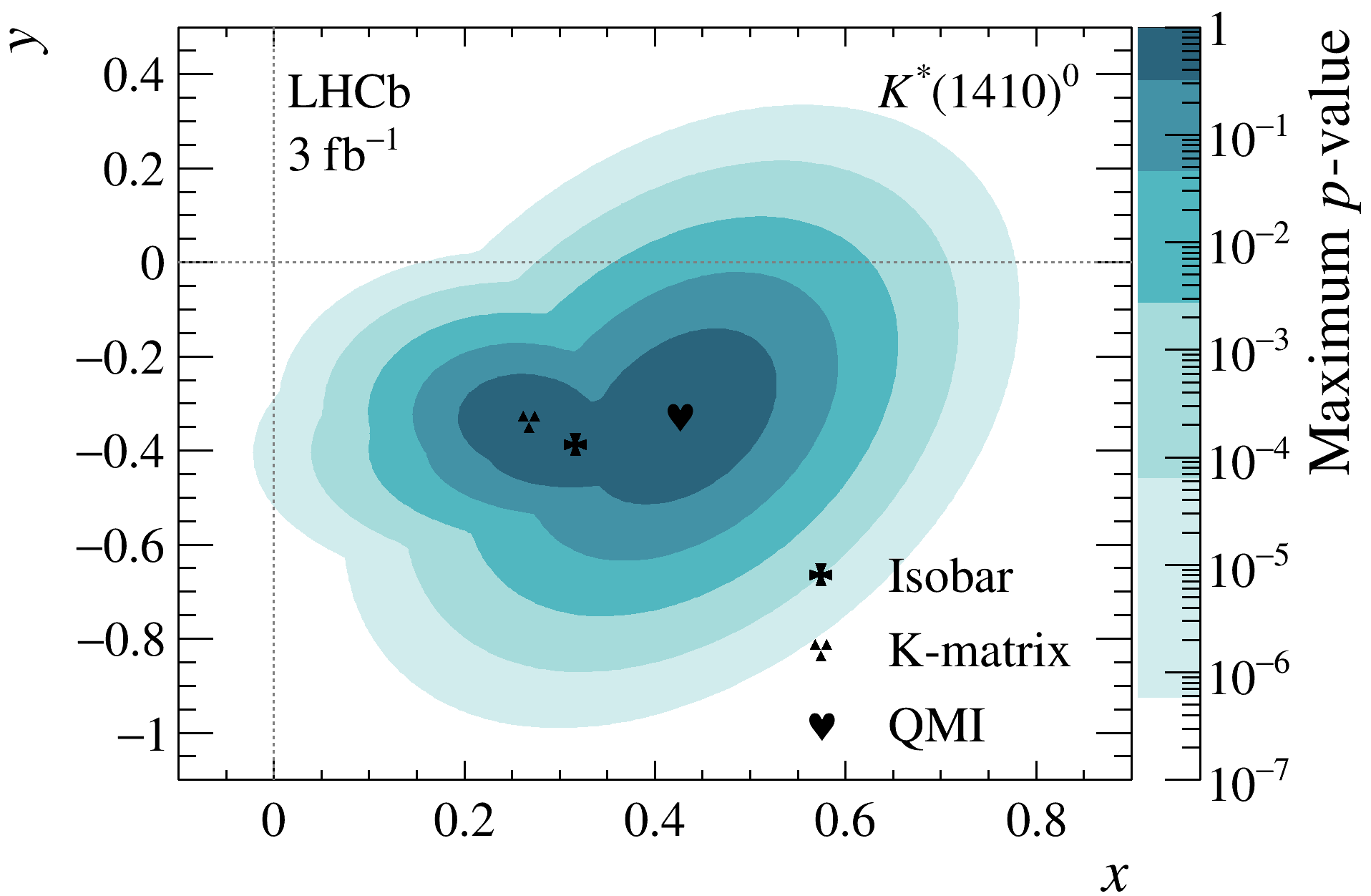}%
    \includegraphics[width=0.5\linewidth]{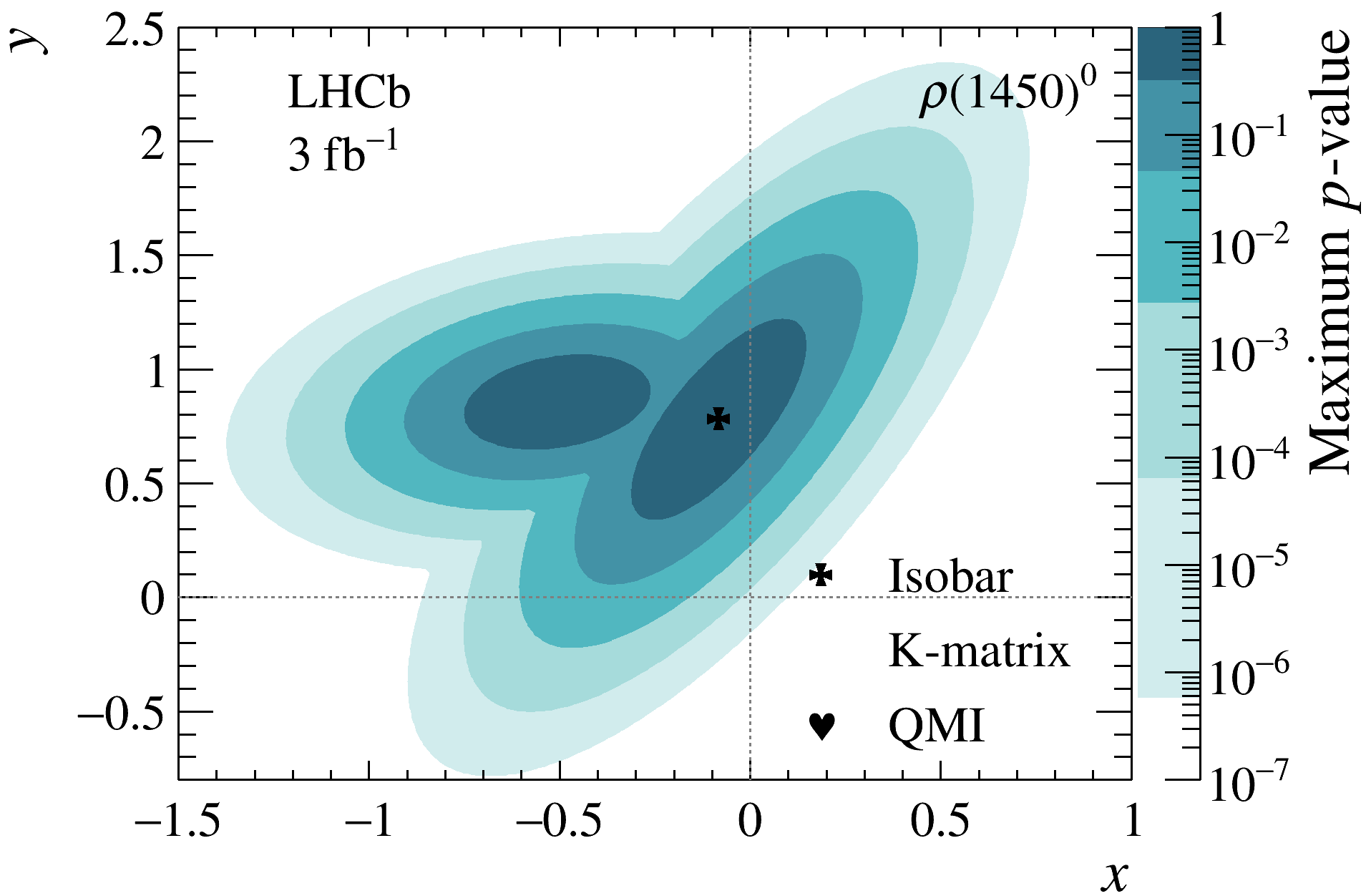}

    \includegraphics[width=0.5\linewidth]{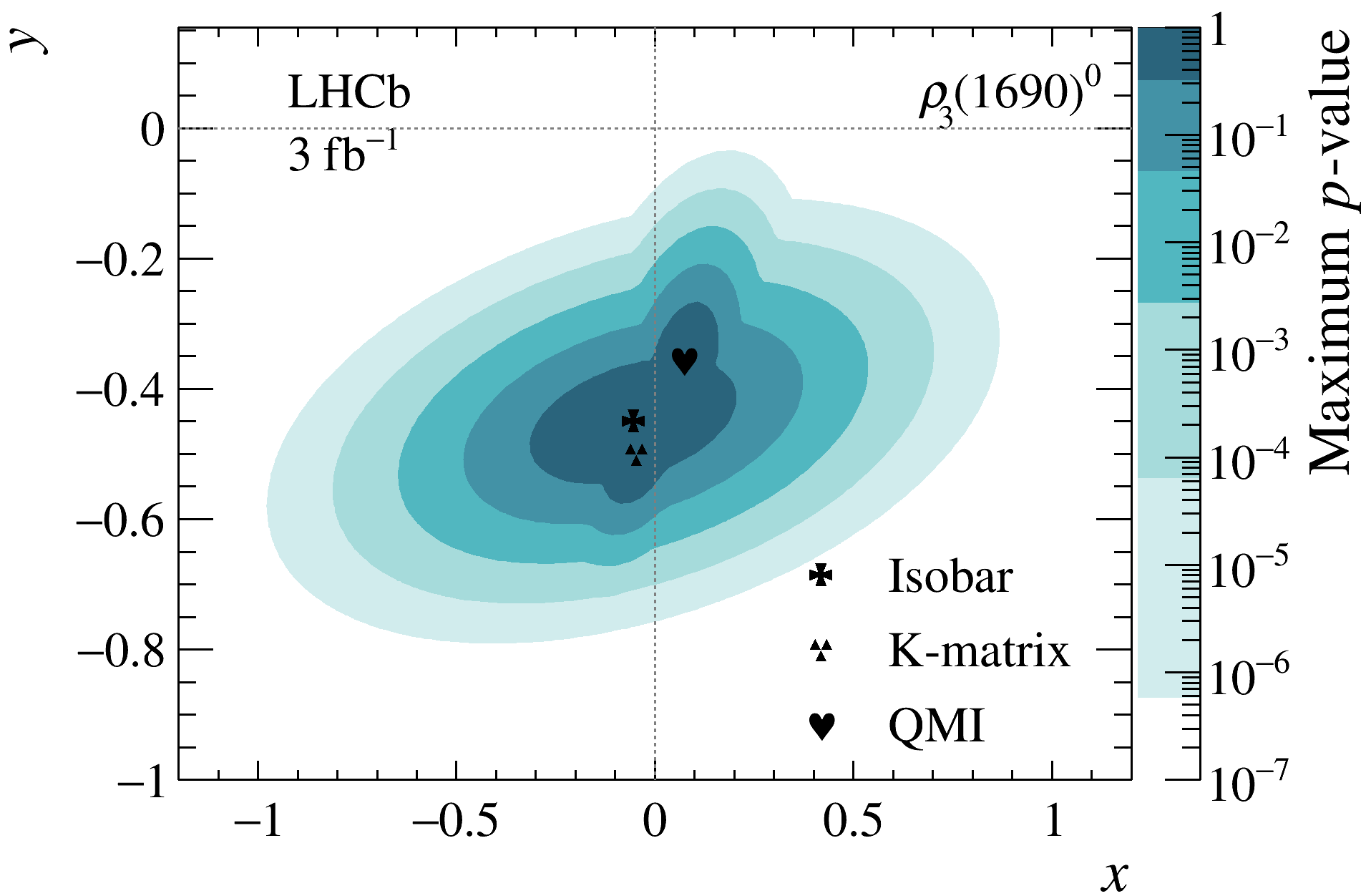}%
    \includegraphics[width=0.5\linewidth]{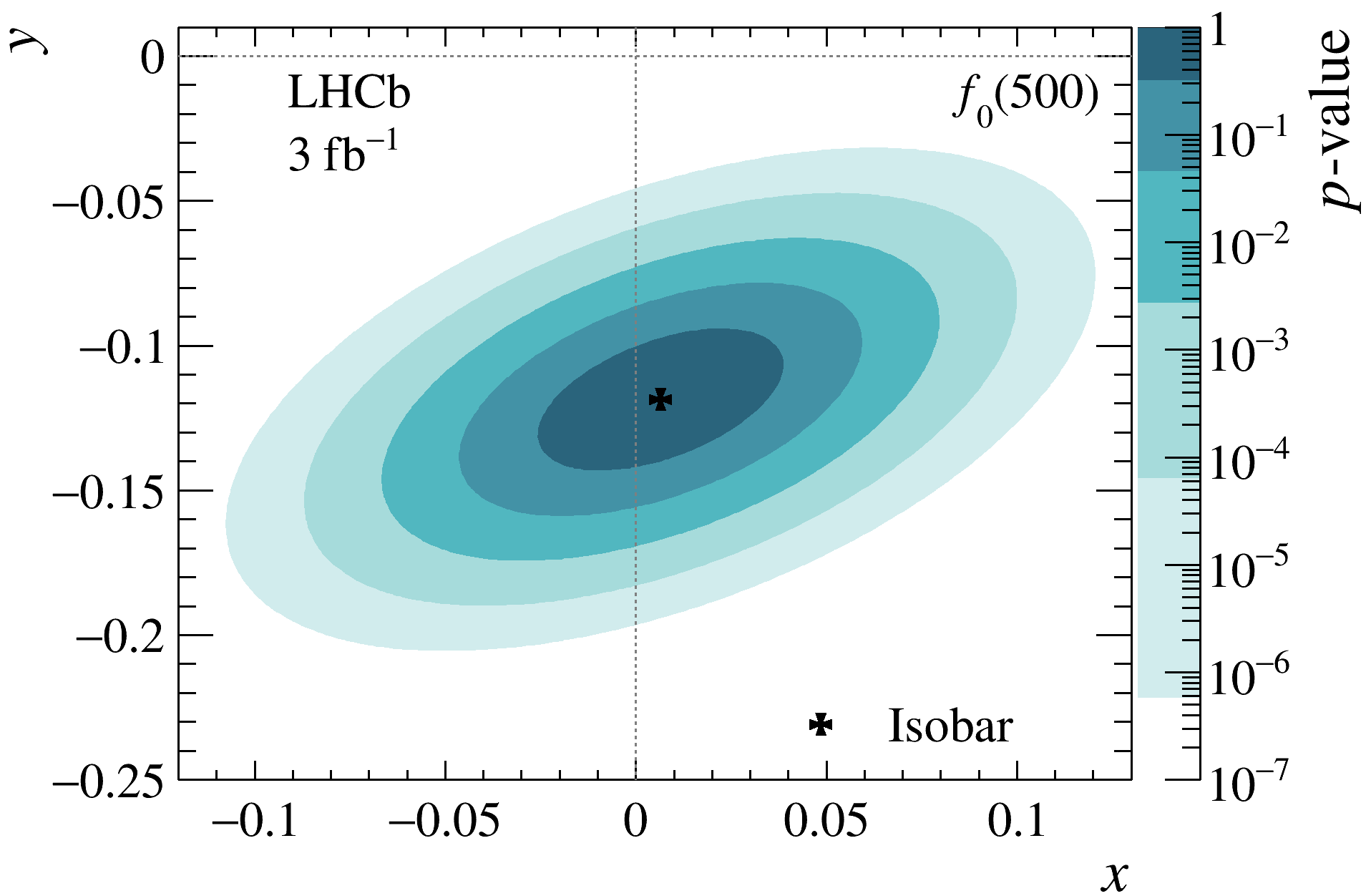}

    \includegraphics[width=0.5\linewidth]{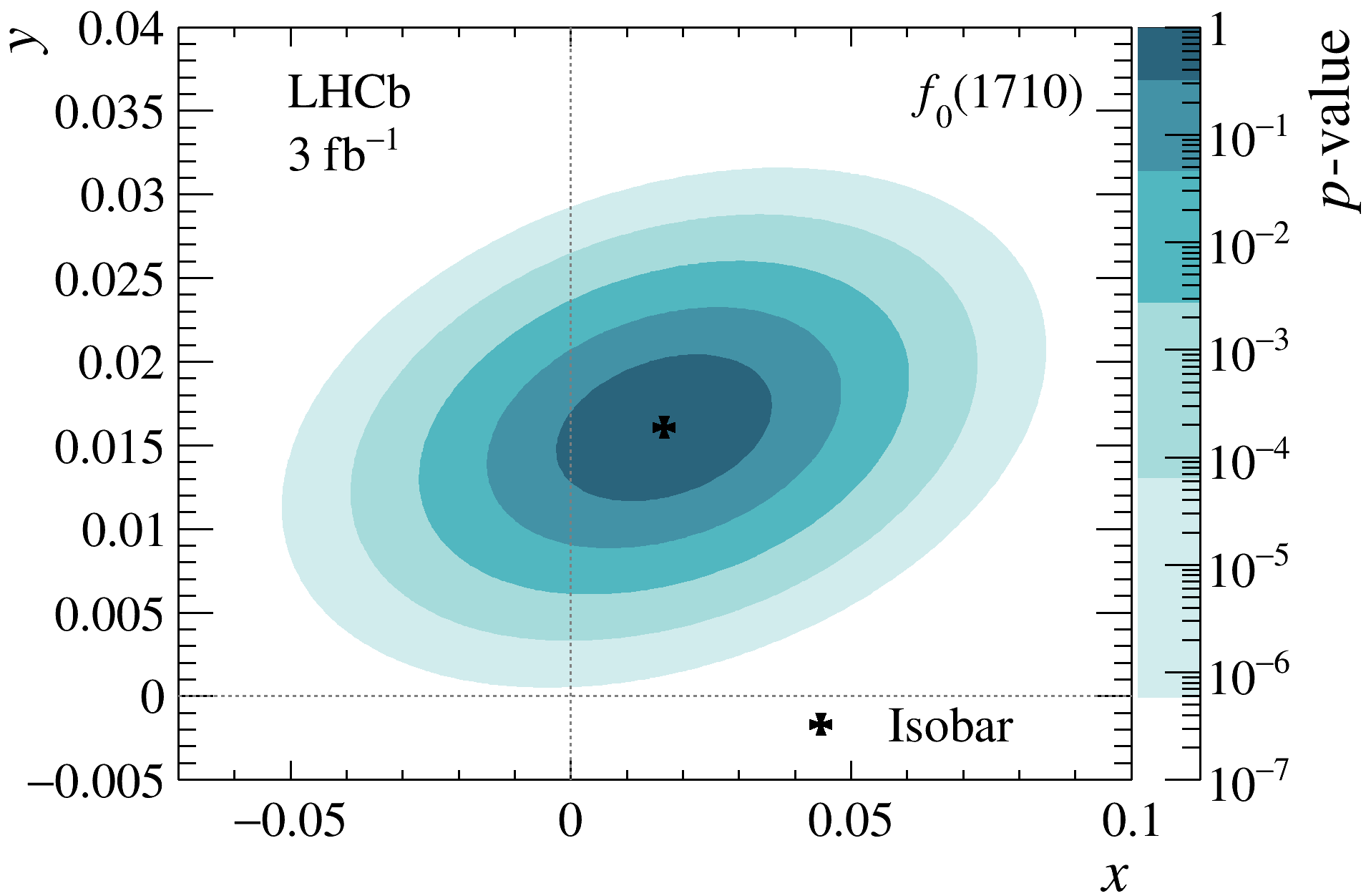}%
    \includegraphics[width=0.5\linewidth]{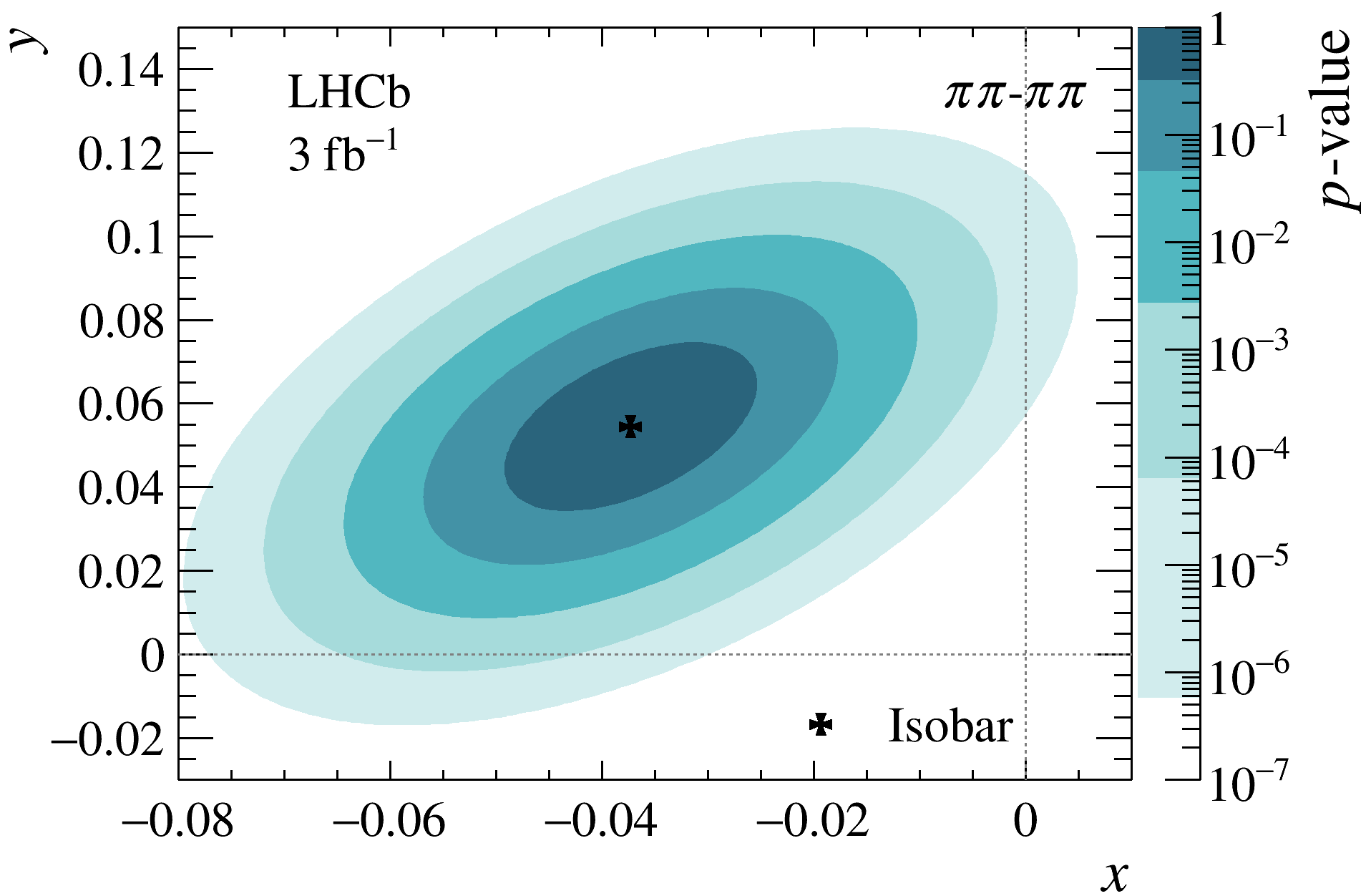}

    \caption{
        Two-dimensional contours of the strong complex coupling for intermediates states listed in the lower half of Table~\ref{tab:significance}. 
        The uncertainties account for statistical, systematic and amplitude-model sources.
    }
    \label{fig:observed}
\end{figure}

For the raw fit-result central values~$\mathbf{\mu}$ and corresponding total covariance matrix~$\mathbf{\Sigma}$ including all systematic and model uncertainties, a parameter-level $\chi^2$ test statistic is constructed as
\begin{equation}
    \chi^2 = (\mathbf{x - \mu})^T \mathbf{\Sigma}^{-1} (\mathbf{x - \mu}) \,.
\end{equation}
Any pair of $(x_i, y_i)$ parameters from $\mathbf{x}$ corresponding to the complex coupling $a_i$ for a given resonance may then be fixed and scanned, with the $\chi^2$ function reminimised at each point. In this way, the potential impact from the remainder of the model through parameter correlations is accounted for.

Adopting a frequentist interpretation similar to that of the \textsc{CKMfitter} group~\cite{CKMfitter2005}, the $\chi^2$ value at each point in the scan is then converted into a $p$-value for two degrees of freedom. 
Confidence intervals can subsequently be derived from the $z$-scores associated with each $p$-value, quantifying the level to which the nonexistence hypothesis of a particular contribution $(x_i, y_i) = (0,0)$, may be rejected with respect to the global minimum. 
This procedure is applied separately to each of the three S-wave approaches, where applicable, with the lowest rejection significance retained for reporting in the main text. 
Significance contours for the newly observed intermediate decays, and that for which first evidence is found, are shown in Fig.~\ref{fig:observed}. 
In these plots, the scans are combined by displaying only the maximum $p$-value of the three S-wave approaches.

\clearpage

\addcontentsline{toc}{section}{References}
\bibliographystyle{LHCb/LHCb}
\bibliography{main,LHCb/standard,LHCb/LHCb-PAPER,LHCb/LHCb-CONF,LHCb/LHCb-DP,LHCb/LHCb-TDR}

\newpage
\centerline
{\large\bf LHCb collaboration}
\begin
{flushleft}
\small
R.~Aaij$^{38}$\lhcborcid{0000-0003-0533-1952},
M.~Abdelfatah$^{69}$,
A.S.W.~Abdelmotteleb$^{57}$\lhcborcid{0000-0001-7905-0542},
C.~Abellan~Beteta$^{51}$\lhcborcid{0009-0009-0869-6798},
F.~Abudin\'en$^{59}$\lhcborcid{0000-0002-6737-3528},
T.~Ackernley$^{61}$\lhcborcid{0000-0002-5951-3498},
A.A.~Adefisoye$^{69}$\lhcborcid{0000-0003-2448-1550},
B.~Adeva$^{47}$\lhcborcid{0000-0001-9756-3712},
M.~Adinolfi$^{55}$\lhcborcid{0000-0002-1326-1264},
P.~Adlarson$^{87}$\lhcborcid{0000-0001-6280-3851}
% ...

{\footnotesize \it

% ...
$^{38}$Nikhef National Institute for Subatomic Physics, Amsterdam, Netherlands\\
% ...
$^{47}$Instituto Galego de F{\'\i}sica de Altas Enerx{\'\i}as (IGFAE), Universidade de Santiago de Compostela, Santiago de Compostela, Spain\\
% ...
$^{51}$Physik-Institut, Universit{\"a}t Z{\"u}rich, Z{\"u}rich, Switzerland\\
% ...
$^{55}$H.H. Wills Physics Laboratory, University of Bristol, Bristol, United Kingdom\\
% ...
$^{57}$Department of Physics, University of Warwick, Coventry, United Kingdom\\
% ...
$^{59}$School of Physics and Astronomy, University of Edinburgh, Edinburgh, United Kingdom\\
$^{60}$School of Physics and Astronomy, University of Glasgow, Glasgow, United Kingdom\\
$^{61}$Oliver Lodge Laboratory, University of Liverpool, Liverpool, United Kingdom\\
% ...
$^{69}$Syracuse University, Syracuse, NY, United States\\
% ...
$^{87}$Department of Physics and Astronomy, Uppsala University, Uppsala, Sweden, associated to $^{60}$\\
% ...
}
\end{flushleft}

\end{document}